\documentclass[aps,prc,twocolumn,showpacs,preprintnumbers,nofootinbib,superscriptaddress,
float,longbibliography]{revtex4-1}

\usepackage{color}
\usepackage[dvipsnames]{xcolor}
\usepackage{graphicx}
\usepackage{dcolumn}
\usepackage{bm}
\usepackage{appendix}
\usepackage{multirow}
\usepackage[colorlinks=true, pdfstartview=FitV, linkcolor=blue,
citecolor=blue, urlcolor=blue]{hyperref}
\usepackage{amsmath}
\usepackage{mathrsfs}

\usepackage{xcolor}
\usepackage[normalem]{ulem}

\def\d{\delta}

\def\be{\begin{equation}}
\def\ee{\end{equation}}
\def\bea{\begin{eqnarray}}
\def\eea{\end{eqnarray}}

\def\kompost{K\o{}MP\o{}ST }
\def\kompostend{K\o{}MP\o{}ST}

\newcommand{\lbk}{\left(}
\newcommand{\rbk}{\right)}

\renewcommand{\vec}{\bm}

\def\nB{f_{_{\rm B}}}
\def\pT{p_{_{\rm T}}}
\def\aem{\alpha_{\rm em}}
\def\Cem{C_{\rm em}}
\def\d{{\rm d}}
\def\lbk{\left(}
\def\rbk{\right)}
\def\sf{{\rm SF}}

\def\ee{{e^+e^-}}
\def\uu{{\mu^+\mu^-}}

\begin{document}


\title{
The polarization of thermal dileptons emitted in high-energy heavy-ion collisions
}

\author{Han Gao}
\affiliation{%
    Department of Physics, McGill University, 
    3600 University Street, Montreal, QC H3A 2T8, Canada}
\author{Xiang-Yu Wu}
 \affiliation{%
   Department of Physics, McGill University, 
   3600 University Street, Montreal, QC H3A 2T8, Canada}
\author{Charles Gale}
 \affiliation{%
   Department of Physics, McGill University, 
   3600 University Street, Montreal, QC H3A 2T8, Canada}
\author{Greg Jackson}
\affiliation{%
   SUBATECH, 
   Nantes Universit\'e, IMT Atlantique, IN2P3/CNRS,
   4 rue Alfred Kastler, La Chantrerie BP 20722, 44307 Nantes, France}
\author{Sangyong Jeon}
 \affiliation{%
   Department of Physics, McGill University, 
   3600 University Street, Montreal, QC H3A 2T8, Canada}
\date{\today}

\begin{abstract}
This work presents calculations of thermal dilepton emission and polarization observables. It features a comprehensive framework which comprises virtual photon spectral functions complete at next-to-leading-order in the strong coupling and iEBE-MUSIC hydrodynamic simulations. The polarization of thermal lepton pairs is shown to be sensitive to  in-medium properties of the quark-gluon plasma. We consider Pb+Pb collisions performed in conditions specific to the LHC and examine  the magnitude and behaviour of the polarization as measured in different frames, the effects of the pre-equilibrium gluon abundance, and we derive a one-to-one mapping between dielectron and dimuon polarization. 
\end{abstract}

\maketitle

%
\section{Introduction}

Relativistic heavy-ion collisions (HICs) can form 
a quark-gluon plasma (QGP): an exotic state of matter described 
by quantum chromodynamics (QCD), the fundamental theory of 
the nuclear   strong interaction. This QGP occupies 
the hot and/or dense regions of the QCD phase diagram and filled 
the universe a few microseconds after the Big Bang. In a HIC event, 
theoretical modeling suggests that the  QGP quickly achieves near 
equilibrium,  evolves hydrodynamically, 
expands, and cools down~\cite{Gale:2013da}. 
As a result of their continual interaction with the medium, 
hadronic observables mostly report on the conditions during 
the late scattering era. In contrast, throughout the evolution history, 
quarks emit electromagnetic (EM) radiation including photons and dileptons. 
As their interaction with the medium is governed by $\alpha_{\rm em}$, 
EM radiation can escape the hot-dense medium almost 
unscathed~\cite{Shuryak:1978ij,Gale:2021emg,Paquet:2015lta,vanHees:2011vb,Gale:2003iz,Gale:1987ki,vanHees:2007th}. 
Therefore, photon and dilepton measurements provide us 
direct information about various properties of the hot medium, 
including electric 
conductivity~\cite{Moore:2006qn,Floerchinger:2021xhb,Rapp:2024grb,Atchison:2024lmf}, 
chemical equilibrium~\cite{Wu:2024pba,Garcia-Montero:2024lbl,Coquet:2021lca,Gale:2021emg}, 
magnetic fields~\cite{Wang:2020dsr,Wang:2021ebh,Wang:2022jxx,Kimura:2024gao} 
and most notably, early-stage 
temperature~\cite{Kajantie:1981wg,NA60:2008ctj,Churchill:2023zkk,Churchill:2023vpt,STAR:2024bpc}. 
Direct photon $\pT$-spectra are effected by the blue-shift owing to 
the hydrodynamical expansion \cite{vanHees:2011vb,Shen:2013vja}. 
Compared with photon emission, dilepton radiation carries 
another kinematic degree of freedom: The invariant mass $M\,$. 
Therefore, dilepton invariant mass spectra, $\d N/\d M$ 
are impervious to flow effects. 
Those two observables, real and virtual photons, 
complement each other well, 
and their measurement in HICs 
has seen significant recent progress~\cite{Scheid:2025gew}. 
\par
At leading order (LO), dilepton radiation is produced by a quark 
and an antiquark annihilating and transforming into a virtual photon.  
The virtual photon with four-momentum $K_\mu$ subsequently decays into 
a lepton-antilepton pair ($\gamma^* \to \ell^+\, \ell^-$) whose 
invariant mass is defined by $M = \sqrt{K^2}=\sqrt{(P_++P_-)^2}$, where $P_+$ and $P_-$ stand for the four-momentum of the $\ell^+$ and $\ell^-$. 
The next-to-leading order (NLO) correction at $O(\aem \alpha_s)$  
involves gluons, which will contribute to Compton scattering 
($g\,q\to \gamma^* \, q$ and $g\,\bar q\to \gamma^* \, \bar q$) 
and modified annihilation processes 
($q \, \bar q \to \gamma^* \, g$ and $q \, \bar q \, g \to \gamma^*$). 
The NLO results will also include the 1-loop corrections to 
$q\,\bar q \to \gamma^*$~\cite{Laine:2013vma,Jackson:2019mop}, 
and the Landau-Pomeranchuck-Migdal effect, 
which is vital for evaluating the dilepton spectrum for general 
$M$~\cite{Arnold:2001ba,Arnold:2001ms,Aurenche:2002pc,Aurenche:2002wq,Ghiglieri:2013gia,Ghiglieri:2014kma}. 
Indeed, previous studies found that the resummed NLO corrections 
significantly enhance the thermal dilepton yields in 
the low-mass region (LMR, $M\lesssim m_\phi$) 
as well as contributing in the 
intermediate-mass region (IMR, $m_\phi \lesssim M \lesssim m_{J/\psi}$)~\cite{Churchill:2023zkk, Churchill:2023vpt,Churchill:2023hog}. 
The high-mass region (HMR, $M \gtrsim m_{J/\psi}$) 
is instead dominated by the Drell-Yan mechanism, 
calculated within the QCD collinear factorization framework. 
\par 
It is also well known that the net thermal dilepton signal in 
the LMR is dominated by processes involving composite 
hadrons~\cite{Rapp:1999ej}. 
In this paper, we focus on thermal dilepton emission in the IMR only, 
where those hadronic reactions are subdominant, 
and where the signal is obtained by calculating 
the spectral functions obtained in thermal field theory with 
partonic degrees of 
freedom~\cite{Ghisoiu:2014mha,Jackson:2019yao,Ghiglieri:2021vcq}. 
\par 
Complementary to the calculation and measurement of their spectra, 
the polarization of dileptons provides insight into the properties 
of the QGP 
medium~\cite{Hoyer:1986pp,Baym:2017qxy,Speranza:2018osi,Wei:2024lah,Seck:2023oyt}. 
The virtual photon spectral function~\cite{Wu:2024vyc} 
can be obtained in perturbation 
theory~\cite{Laine:2013vma,Jackson:2019mop,Jackson:2022fqj,Churchill:2023vpt} 
and studied on the 
lattice~\cite{Ghiglieri:2016tvj,Ce:2020tmx,Ce:2022fot,Ali:2024xae}. 
Plasma momentum anisotropy influences the polarization of both 
dileptons~\cite{Coquet:2023wjk} and real photons~\cite{Hauksson:2023dwh}, 
and external electromagnetic fields can also change the dilepton 
polarization signature~\cite{Wei:2024lah}. 
The angular distribution of lepton pairs therefore contains precious 
information related to the very nature of 
the medium from which they are emitted. 
\par 
In this work, we perform a detailed exploration of dilepton polarization 
at NLO, in view of eventual measurements performed 
in relativistic HIC experiments. 
We concentrate  on conditions (colliding systems and energy) 
occuring at the LHC. 
We start with a comprehensive, Lorentz covariant description of 
both dilepton production and dilepton polarization in terms of 
the photon spectral function in Sec.~\ref{sect:theory}. 
The method of predicting dilepton polarization signals with 
a multi-stage hydrodynamic background and a phenomenological approach 
to modeling the pre-equilibrium dilepton production is introduced 
in Sec.~\ref{sect:hydro}. In Sec.~\ref{sect:results}, 
we present and discuss our results for different frames, 
different pre-equilibrium modeling, and connect 
the polarization of dielectrons to that of dimuons. 
Finally, a  summary is given in Sec.~\ref{sect:summary}. 

%
\section{Thermal dilepton production and polarization}\label{sect:theory}

\subsection{The kinematics of the dilepton polarization}

The dilepton production rate (DPR) in a finite-temperature medium 
defines the number of lepton-antilepton pairs 
produced from a unit four-volume. 
For a uniform plasma in its fluid rest frame, 
the fully differential DPR $R_{\ell \bar \ell}$ 
is given by~\cite{Kapusta:2023eix}
\begin{equation}\label{eq:d6rate}
  E_+ E_- 
  \frac{
    {\rm d} R_{\ell \bar \ell}
  }{
    {\rm d}^3 \vec p_+^{ }
    {\rm d}^3 \vec p_-^{ }
  } 
  \; = \; 
  -\frac{2e^4 \; C_{\rm em}}{(2\pi)^6} \,
  \frac{\nB(\omega)}{(K^2)^2} \,
  L^{\mu\nu} \,
  \rho_{\mu\nu}(\omega,|\vec k|) \,
  \; .
\end{equation}
Here, $C_{\rm em} = \sum_{f=u,d,s}Q_f^2 = 2/3$ is the 
sum of the squared quark charges 
(in units of $e$) for a 3-flavour QCD plasma, 
$\nB(\omega) = [\exp(\omega/T)-1]^{-1}$ is the Bose-Einstein distribution, 
$K = P_+ + P_-$ is the four-momentum of 
the virtual photon from which the leptons decay. 
The photon spectral function $\rho^{\mu\nu}(\omega,|\vec k|)$ 
is a function of the dilepton energy $\omega$ and momentum magnitude 
$|\vec k|$ in the fluid rest frame, and involves 
the temperature $T$ and the baryon chemical potential $\mu_B$. 
The leptonic tensor $L_{\mu\nu}$ follows  from the final state 
spin sum and is 
\begin{equation}
  L_{ }^{\mu\nu} 
  \; \equiv \;
  P_+^\mu P_-^\nu + P_-^\mu P_+^\nu 
  - g_{ }^{\mu\nu} 
  \big( P^{ }_+ \cdot P^{ }_- + m_\ell^2 \big)
  \; .
\end{equation}
\par
We will label the frame where the lepton pairs are measured, 
``the lab frame''. 
Each fluid cell in the fireball  has its own local rest frame: 
one  Lorentz-transforms from  the lab-frame $K_{\rm lab}$ 
to the local rest frame of each fluid cell using 
$K_{\rm lrf}^\mu = \Lambda^\mu_\nu(u)K_{\rm lab}^\nu$ 
where $\Lambda^\mu_\nu(u)$ is the Lorentz transform matrix, 
i.e. $(1,\vec 0)^\mu = \Lambda^\mu_\nu(u)u^\nu$. 
Alternatively, one may first generalize Eq.~\eqref{eq:d6rate} to an 
arbitrary frame where the fluid velocity is $u^\mu = \gamma(1,\vec v)$. 
Noting that in the local rest frame, the DPR given by 
Eq.~\eqref{eq:d6rate} only depends on the dilepton 
momentum $K$ via $\omega$ and $|\vec k|\,$. 
Both can be generalized to any frame with $\omega \to u\cdot K$ 
and $|\vec k| \to \sqrt{(u\cdot K)^2 - K^2}$~\cite{Weldon:1982aq}. 
Then by using 
$
  \frac{\d^3\vec p_\pm}{2E_\pm} 
  = 
  \delta(P_\pm^2 - m_\ell^2) \d^4 P^{ }_{\pm} 
$, 
we can rewrite Eq.~\eqref{eq:d6rate} into a manifestly covariant form,
\begin{eqnarray}\label{eq:d8rate}
  \frac{
    {\rm d} R_{\ell \bar \ell}
  }{
    {\rm d}^4 P_+^{ } 
    {\rm d}^4 P_-^{ }
  } 
  & = & 
  \; - \;
  \frac{ e^4 \; C_{\rm em} }{ 2(2\pi)^6 } \, 
  \frac{\nB(u\cdot K)}{K^4} \,
  L^{\mu\nu} \,
  \rho_{\mu\nu}
  \nonumber\\
  & \times &
  \delta(P_+^2 - m_\ell^2) \,
  \delta(P_-^2 - m_\ell^2) \,
  \; .
\end{eqnarray}
Inserting $1=\int \d^4 K \, \delta^{(4)}\left(K - P_+ - P_- \right)$ 
and integrating over the lepton four-momenta 
gives rise to the standard expression for the DPR in terms of 
$K$~\cite{McLerran:1984ay,Weldon:1990iw,Gale:1990pn}:
\begin{eqnarray}\label{eq:d4rate}
  \frac{\d R_{\ell\bar\ell}}{\d^4 K} 
  & = & 
  \frac{ \alpha_{\rm em}^2}{3\pi^3} 
  \ \frac{C_{\rm em}}{K^2}
  \, B\bigg(\frac{m_\ell^2}{K^2}\bigg) \,
  \rho^{ }_{_{\rm V}} 
  \, \nB(u\cdot K)
  \; , \\[-2mm]
  \nonumber
\end{eqnarray}
where $\rho_{_{\rm V}} \equiv \rho^{\mu}_\mu$ 
is the ``vector channel'' spectral function and 
$B(\xi) \equiv (1+2\xi)\sqrt{1-4\xi}$ 
is a kinematic phase-space factor.\footnote{%
  $B(\xi) = 0$ for $\xi>1/4$, 
  which states the decay 
  $\gamma^*\to\ell^+\ell^-$ is kinematically forbidden if $M<2m_\ell$.
}

In equation~\eqref{eq:d4rate}, the angular distribution of 
the lepton momentum has been integrated out. 
However, the dilepton polarization will reveal the spin polarization of 
the virtual photon, 
which subsequently provides information about medium properties. 
In an environment where a fluid velocity $u^\mu$ exists, 
a general expansion of the tensorial spectral function is\footnote{%
  This decomposition, Eq.~\eqref{eq:decom}, will no longer be general 
  if any other vector or tensor elements are also present, 
  for example, a vorticity or a background electromagnetic field. 
  These elements, although naturally present for a QGP created by HIC, 
  are beyond the scope of this study.
}

\begin{eqnarray}
\label{eq:decom}
  \rho^{\mu\nu} 
  & = & 
  \rho_{_{\rm T}}  
  \left( g^{\mu\nu}  - \frac{K^\mu K^\nu}{K^2}\right) 
  \nonumber \\[2mm]
  & - & \!
  \frac{
    \raisebox{1mm}{$\displaystyle \rho_{_\Delta}$}
  }{
    \raisebox{-3mm}{$\displaystyle 
      1 - \displaystyle\frac{(u\cdot K)^2}{K^2}
    $}}
  \left( u^\mu - \frac{u\cdot K}{K^2}K^\mu\right) \!
  \left( u^\nu - \frac{u\cdot K}{K^2}K^\nu\right) 
  \nonumber\\[-4mm]
\end{eqnarray}
where the Ward identity $K_\mu \rho^{\mu\nu} = 0$ is respected. 
The transverse and longitudinal parts of the spectral function 
are defined as 
\begin{equation}
  \label{eq:rho_T_L}
  \rho^{ }_{_{\rm T}} 
  \; \equiv \;
  \frac{\rho^\mu_\mu - \rho^{ }_{_{\rm L}}}2 
  \,,\quad 
  \rho^{ }_{_{\rm L}} 
  \; \equiv \; 
  -\frac{K^2 }{(u\cdot K)^2 - K^2} \; 
  \rho^{ }_{\mu\nu}u^\mu u^\nu
  \,,\quad 
\end{equation}
while 
$\rho_{_\Delta} = \rho_{_{\rm T}} - \rho_{_{\rm L}}$ 
stands for the difference of the transverse and the longitudinal part. 
\par 
Returning to the final state kinematics of the lepton pair, 
as relevant for polarization, 
we define the dilepton momentum difference $l^\mu$ as 
\begin{equation}\label{eq:ellmu}
  l^\mu \; = \; P_+^\mu - P_-^\mu 
\end{equation}
so that $P_\pm^\mu  = K^\mu  \pm l^\mu/2\,$. 
Substituting Eqs.~\eqref{eq:decom}, \eqref{eq:ellmu} 
and the expression of $L^{\mu\nu}$ into Eq.~\eqref{eq:d8rate}, 
we may integrate over the magnitude $l = | \vec{l} |$ 
to obtain the DPR as a function of the relative angular 
distribution of the final leptons. 
Doing so, we find
\begin{widetext}
\begin{eqnarray}\label{eq:drdkdo}
  \frac{{\rm d}R_{\ell\bar\ell}}{{\rm d}^4K\d\Omega_\ell} 
  & = & 
  \int_{-\infty}^{\infty} \d l^0 
  \int_0^\infty \d |\vec l| \; |\vec l|^2
  \frac{{\rm d}R_{\ell \bar \ell}}{{\rm d}^4K\d^4l} 
  \\[2mm]
  & = & 
  \frac{ \aem^2C_{\rm em} }{
    {8\pi}^3
  }  \;
  \frac{\nB(u\cdot K)}{ b^{3/2} K^2} \;
  \frac{\sqrt{K^2 - 4 m_\ell^2}}{u\cdot K} 
  \Bigg[ \, 
  2\left(1+\frac{2m_\ell^2}{K^2}\right) 
  \rho_{_{\rm T}}
  - 
  \bigg( 
    u_0\frac{\vec k\cdot \hat l}{u\cdot K} 
    - 
    \vec u \cdot \hat l
  \bigg)^{\!2}
  \left( 
    1-\frac{4m_\ell^2}{K^2}
  \right)
  \frac{ \rho_{_\Delta} }{ a b }
  -\rho_{_\Delta} 
  \, \Bigg]
  \; ,
  \nonumber
\end{eqnarray}
\end{widetext}
where $\hat l = \vec l /|\vec l|\,$ 
and
\begin{equation}\label{eq:defab}
    a 
    \; \equiv \; 
    1 - \frac{(u\cdot K)^2}{K^2}
    \; ,
    \qquad 
    b 
    \; \equiv \; 
    1 - \frac{( \vec k\cdot \hat l)^2}{(u\cdot K)^2}
    \; .
\end{equation}

\subsection{Polarization coefficients}
\label{sect:pol_coeffs}

Equation~\eqref{eq:drdkdo} 
describes the distribution of the direction 
of the dilepton momentum difference $\vec l$ 
for a given dilepton pair of 4-momentum $K$, 
and it holds in any reference frame. 
However, this equation can be simplified further if we choose 
to measure $\hat l$ in 
a comoving frame of the virtual photon ($\vec k = 0$): 
\begin{widetext}
\begin{equation}\label{eq:genpol}
  \frac{ \d R_{\ell\bar\ell} }{ \d^4K\d\Omega_\ell } 
  \; = \;
  \frac{\aem^2 \Cem}{{8\pi}^3} \;
  \frac{ \nB(u\cdot K) }{ K^2 } \;
  B\left(\frac{m_\ell^2}{K^2}\right) 
  \bigg[ \;
  2\left( 1 + \frac{2m_\ell^2}{K^2} \right) \rho_{_{\rm T}} +
  \frac{(\vec u_*\cdot \hat l)^2}{\vec u_*^2}
  \left( 1-\frac{4m_l^2}{K^2}\right) \rho_\Delta - \rho_\Delta 
  \;\bigg]
  \; .
\end{equation}
\end{widetext}
Here, 
we use $\vec u_*$ to denote the fluid velocity in the rest frame of 
the virtual photon $\gamma^*$, 
which is in general different from the fluid velocity of the lab frame 
denoted hereon by  $\vec u\,$. 
For a given lab-frame virtual photon four-momentum 
$K^\mu = (\omega,\vec k)$, 
\begin{equation}
  \label{eq:ustar}
  \vec u_* 
  \; = \;
  \vec u 
  \, + \,
  \left( \frac \omega M - 1\right) (\vec u \cdot \hat k)\hat k 
  \, - \, 
  \frac{u_0 \vec k}{M}.
\end{equation}
\par 
Dilepton polarization coefficients are defined by the following expansion 
of the differential yields 
$\frac{\d N_{\ell\bar\ell}}{\d^4K\d\Omega_\ell}$ 
in the rest frame of the virtual photon 
(with a defined lab-frame four-momentum $K^\mu$)
\begin{eqnarray}\label{eq:defpol}
   \frac{{\rm d}N_{\ell\bar\ell}}{{\rm d}^4 K \, {\rm d}\Omega_\ell } 
   & = & 
   {\cal N} \; \bigg(
   1 
   +
   \lambda_\theta \cos^2 \theta_\ell 
   +
   \lambda_\phi \sin^2\theta_\ell \cos 2\phi_\ell 
   \nonumber \\[1mm]
   & + & 
   \lambda_{\theta\phi} \sin 2\theta_\ell \cos \phi_\ell 
   \hspace{2mm} +
   \lambda_\phi^\perp \sin^2\theta_\ell \sin 2\phi_\ell 
   \nonumber \\[1mm]
   & + & 
   \lambda^\perp_{\theta \phi} \sin2\theta_\ell \sin \phi_\ell \bigg)
   \; ,
\end{eqnarray}
where the angles $\theta_\ell$ and $\phi_\ell$ are defined by 
the direction of the dilepton momentum difference 
$
  \hat l 
  = 
  (\sin\theta_\ell\cos\phi_\ell,
  \sin\theta_\ell \sin \phi_\ell,
  \cos\theta_\ell)\,
$.  
It is evident from Eq.~\eqref{eq:genpol} that 
the thermal dilepton polarization is induced by the 
$(\vec u \cdot \hat l)^2$ term, 
which confirms our previous intuition that a special direction set by 
the fluid velocity would give rise to 
the anisotropic angular distribution of the decayed leptons. 
Expanding the $(\vec u_* \cdot \hat l)^2$ term and 
comparing Eqs.~\eqref{eq:genpol} and \eqref{eq:defpol}\footnote{%
  For a uniform plasma that we are now considering, 
  $\d N$ and $\d R$ only differ by an overall space-time volume factor. 
} 
gives the following explicit representations of the 
individual polarization coefficients:
\begin{eqnarray}
  \lambda^{ }_{\theta} \label{eq:ltheta}
  & = & 
  \frac{
     3\lbk \chi_z-\frac 1 3\rbk \lbk 1- 4\xi\rbk \rho^{ }_\Delta
  }{D}
  \; ; \\[1mm]
  \lambda^{ }_\phi \label{eq:lphi}
  & = &
  \frac{
    \lbk \chi_x - \chi_y \rbk\left(1 -4\xi\right)\rho^{ }_\Delta
  }{D} 
  \; ; \\[1mm]
  \lambda^{ }_{\theta\phi} \label{eq:lthph}
  & = &
  \frac{
    2\chi_{xz} \lbk 1- 4\xi\rbk \rho^{ }_\Delta
  }{D} 
  \; ; \\[1mm]
  \lambda_\phi^\perp
  & = &
  \frac{
    2\chi_{xy}  \lbk 1- 4\xi\rbk \rho^{ }_\Delta
  }{D}
  \; ; \\[1mm]
  \lambda^\perp_{\theta \phi} \label{eq:lperpthph}
  & = &
 \frac{
   2\chi_{yz} \lbk 1- 4\xi\rbk \rho^{ }_\Delta
 }{D} 
 \; .
\end{eqnarray}
Here we have introduced the short-hand notations 
$\chi_i \equiv (u_*^i)^2/\vec u_*^2$, 
$\chi_{ij} \equiv u_*^i u_*^j/\vec u_*^2$ 
and 
$\xi \equiv m_\ell^2/K^2$. 
The common denominator is 
$
  D
  \equiv
  \frac{4}{3} \lbk 1+2\xi\rbk \rho_{\rm V} -
  \lbk \chi_z- \frac{1}{3}\rbk \lbk 1- 4\xi\rbk \rho_\Delta\,
$. 
Note that the five polarization coefficients are not independent, 
as they are uniquely determined by three degrees of 
freedom discussed in Appendix~\ref{app:dof}.

It should be clear from Eqs.~\eqref{eq:ltheta}--\eqref{eq:lperpthph}, 
that the spectral function $\rho^{ }_\Delta$
(defined below Eq.~\eqref{eq:rho_T_L}) 
plays a key role in describing polarization properties of 
dilepton spectra.\footnote{%
  The notation $\rho_{\rm H} = 2 \rho_{\Delta}$ 
  is frequently used in the literature. 
} 
We mention that this spectral function has several properties 
which make it of broader theoretical interest. 
Firstly, it vanishes identically in vacuum and is 
highly suppressed for large $M$~\cite{Brandt:2017vgl}. 
The leading behaviour for $\omega,k \gg T$ can be obtained 
using an 
Operator Product Expansion~\cite{Caron-Huot:2009ypo,Laine:2010tc}, 
\begin{eqnarray}
  \rho^{ }_{\rm V}
  & = &
  \bigg( 1 + \frac{\alpha_s}{\pi} \bigg)
  \frac{3 M^2}{4\pi^3}
  \, + \, 
  O\bigg( \frac{T^4 \omega^2}{M^4}, \frac{T^4 k^2}{M^4} \bigg)
  \; , 
  \\[1mm]
  \rho^{ }_\Delta
  & = &
  \alpha_s \frac{32 \pi^2 k^2 T^4 }{27 M^4} 
  \, + \, 
  O\bigg( \frac{T^6}{M^4} \bigg) 
  \; .
\end{eqnarray}
The spectral functions also satisfy 
general sum rules~\cite{Gubler:2017qbs}, 
which for $\rho^{ }_\Delta$ reads 
$\int_0^\infty {\rm d} \omega \, \omega \rho^{ }_\Delta = 0\,$.
This has made $\rho^{ }_\Delta$ particularly 
conducive to lattice 
studies~\cite{Ali:2024xae,Ce:2022fot,Ce:2020tmx,Brandt:2017vgl}. 
However, so far these investigations have focused on photon 
production (proportional to $\rho^{ }_\Delta(k,k)$), whereas 
rigorous constraints on dileptons will benefit from 
simultaneous extraction of 
$\rho^{ }_\Delta$ and $\rho^{ }_{\rm V}\,$~\cite{Meyer:2023ntn}. 
Nevertheless, current information about $\rho^{ }_\Delta$ 
gives a qualitative picture of the dilepton polarisation 
coefficients in Eqs.~\eqref{eq:ltheta}--\eqref{eq:lperpthph} 
($\rho^{ }_{\rm V}$ is important for the normalisation, 
due to the denominator $D$ which is mainly set by 
the ultraviolet $\sim M^2$ piece). 

\subsection{The Helicity frame, the Collins-Soper frame,\\[1mm] and the rotationally-invariant $\tilde\lambda$}

%
\begin{figure}[ht]
\centering
\includegraphics[width=1\linewidth]{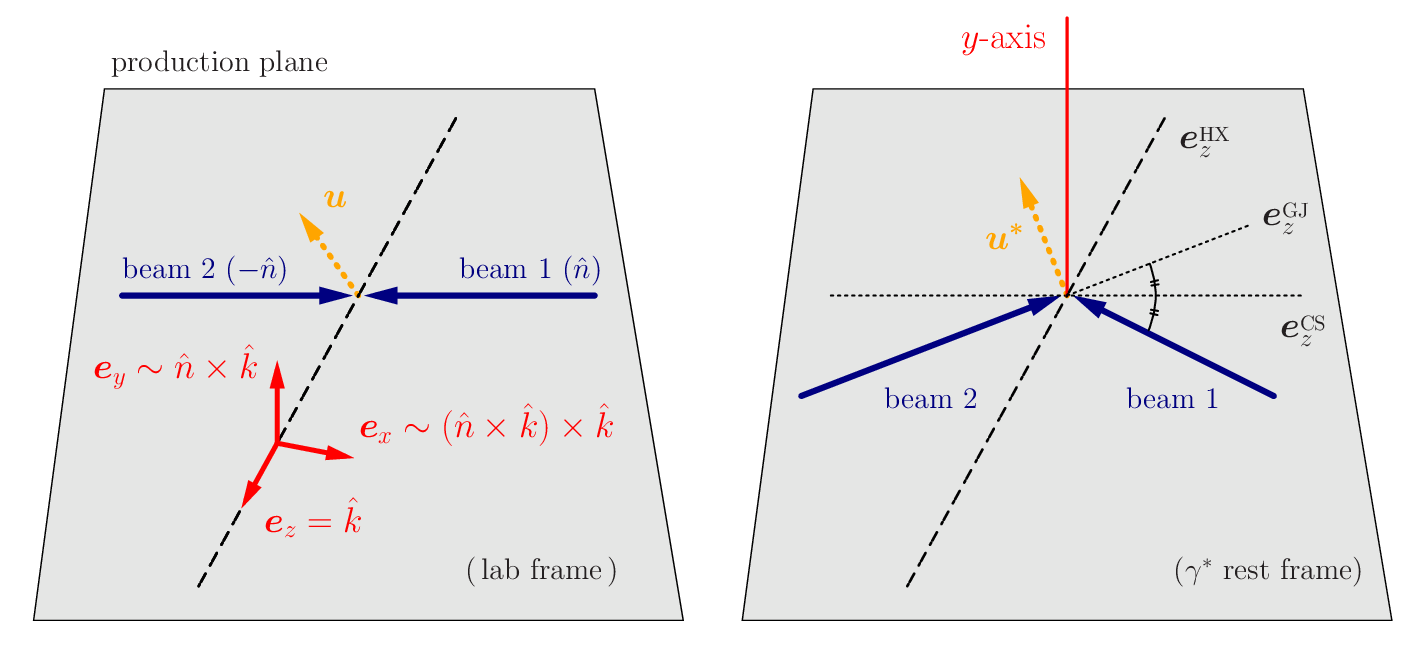}
\vspace{-2mm}
\caption{
  Illustration of the helicity (HX), Collins–Soper (CS), 
  and Gottfried–Jackson (GJ) frames. 
  The left panel is in the laboratory frame, while the right panel 
  corresponds to the local rest frame of the virtual photon. 
  The production plane is defined by the momenta of the 
  two colliding beams and the virtual photon. 
  In  the HX frame, the $\hat e_z$-axis is chosen to be parallel 
  to the momentum of the virtual photon. 
  In  the CS frame, the $\hat e_z$-axis is defined as the bisector of 
  the angle between the two beam directions. 
  In the Gottfried-Jackson (GJ) frame, the $\hat e_z$-axis is 
  taken along the momentum of one of the two colliding beams~\cite{Faccioli:2010kd}.
}     
\label{fig:frames}
\end{figure}
%

We have yet to specify axes in the virtual photon rest frame. 
A common choice is the helicity frame (HX), 
where $z$ direction is defined by $\hat e_z^{\rm HX} = \hat k$ in the lab frame, 
and $\hat e_y^{\rm HX} = \hat n \times \hat k/|\hat n \times \hat k|$ where 
$\hat n$ stands for the beam direction of the initial collision. 
In HX frame, we note that $\lambda_\theta$ is the only 
non-vanishing polarization coefficient for a medium at rest,
$\vec u =  0$. 
In that case, Eq.~\eqref{eq:ustar} reads 
$\vec u_* = -\vec k /M$, giving 
$\chi_z = 1, \chi_x = \chi_y = 0$. 
With the approximation $m_\ell \simeq 0$, 
Eq.~\eqref{eq:ltheta} becomes 
\begin{equation}\label{eq:lth_approx}
    \lambda_\theta^{\rm HX} \Big|_{\vec{u}=0}
    \; \simeq \; 
    \frac{
      \rho_\Delta
    }{
      \rho_{_{\rm T}}  + \rho_{_{\rm L}} 
    }
    \; .
\end{equation}
As the only coefficient that is non-vanishing 
for a  medium at rest, $\lambda_\theta$ measurements 
in HX frame would be a good and simple 
indicator of $\rho_\Delta \equiv \rho_{_{\rm T}}  - \rho_{_{\rm L}}\,$. 
\par
Another popular choice is the 
Collins-Soper (CS) frame~\cite{Collins:1977iv}. 
The CS frame has its $z$-axis to be the bisector of the angle between 
the momentum of one initial ion and the opposite of that of the other ion 
(in the virtual photon rest frame). 
One can work out the analytical expression of the CS-frame 
$z$-direction in terms of the beam direction $\hat n$ 
and the virtual photon direction $\hat k$ in the lab frame, to find
\begin{equation}\label{eq:ezcs}
  \vec e_z^{\rm CS} 
  \; = \;
  \frac{
    \omega \hat n - (\omega - M)(\hat n \cdot \hat k)\hat k
  }{
    \sqrt{\omega^2 - (\hat n \cdot \vec k)^2}
  }
  \; .
\end{equation}
The CS frame and the HX frame are related by a rotation 
along their common $y$ axis~\cite{Speranza:2018osi}, 
illustrated in Fig.~\ref{fig:frames}. 

A combination of the polarization coefficient which is invariant under such a rotation~\cite{Faccioli:2010ji} is given by
\begin{eqnarray}\label{eq:invlambda}
        \widetilde \lambda 
        & \equiv & 
        \frac{
          \lambda_\theta + 3\lambda_\phi
        }{
          1 - \lambda_\phi
        } \\[1mm] 
        & = & 
        \frac{
          3\lbk \chi_x+\chi_z-\chi_y-\frac 1 3\rbk 
          \lbk 1- 4\xi\rbk 
          \rho_\Delta
        }{
          \frac{4}{3} \lbk 1+ 2\xi\rbk \rho_{\rm V} 
          -
          \lbk \chi_x+\chi_z-\chi_y-\frac 1 3\rbk
          \lbk 1- 4\xi \rbk \rho_\Delta
        }
        \nonumber \; ,
\end{eqnarray}
as $\chi_x + \chi_z = \left[(u_*^x)^2 + (u_*^z)^2\right]/\vec u_*^2$ 
is an invariant with respect to rotations about the $y$-axis. 

\subsection{Superposition of polarization coefficients}
\label{sect:superp_coeff}

We have so far limited our discussion to  
the dilepton polarization from a uniform plasma. 
In general, the system created by a heavy-ion collision is 
highly dynamic but 
the QGP can be treated as a collection of fluid cells within 
a hydrodynamic description. 
For each cell, the DPR and polarization are calculated under 
the assumption of local equilibrium. 
The dilepton differential 
\emph{yield} for a given lab-frame virtual photon four-momentum $K$ 
is given by an integration over 
the entire history of the QGP evolution and all fluid cells, as
\begin{equation}\label{eq:yield}
  \frac{\d N_{\ell \bar \ell}}{ \d^4 K } 
  \; = \; 
  \int \d^4 X 
  \ \frac{\d R_{\ell \bar \ell}}{\d^4 K}
  {
  \Big( T(X), \mu_{_{\rm B}}(X), \vec{u}(X), ... \Big)
  }
  \; .
\end{equation}
Here, as discussed after Eq.~\eqref{eq:d4rate}, 
a Lorentz boost from the lab frame to
the local rest frame is preformed for every fluid cell 
(located at the space-time point $X=(t,\mathbf{x})$) 
in order to evaluate the spectral function. 
\par 
The situation becomes more complicated when discussing polarization. 
For a specified lab-frame virtual photon momentum $K$ and 
for each fluid cell, two important reference frames need to be considered: 
(1) the fluid rest frame 
where the spectral function is evaluated, 
and (2) the virtual photon's rest frame, 
where the fluid velocity is given by Eq.~\eqref{eq:ustar}. 
Taking into account the relevant transformations needed, 
the final observed polarization coefficients 
are given by a weighted average, 
obtained by superposing both sides of Eq.~\eqref{eq:defpol} 
for all fluid cells 
\begin{equation}\label{eq:avg_lambda}
    \lambda^{ }_i(K) 
    \; = \; 
    \frac{
    \raisebox{3mm}{$
      \displaystyle
      \int \d^4 X 
      \ \lambda^{ }_i
      \Big( K;T(X),\vec u(X) \Big) \;
      {\cal N}\big(K, X \big)
      $}
    }{
    \raisebox{-3mm}{$
      \displaystyle
      \int \d^4 X\  
      {\cal N}\big( K, X \big)
      $}
    }
\end{equation}
for $i = \{ \theta,\phi,\theta \phi,\dots \}$, 
with the weight function ${\cal N}(K,X)$ 
relating to the local DPR by integrating over 
$\d \Omega_\ell$ on both sides of Eq.~\eqref{eq:defpol}, 
explicitly given by 
\begin{equation}\label{eq:defN}
    {\cal N}
    \big(K, X \big) 
    \; = \; 
    \frac{
    \raisebox{3mm}{$
      \displaystyle
      \frac{\d R_{\ell \bar \ell}}{\d^4 K}
      {
      \Big( T(X), \mu_{_{\rm B}}(X), \vec{u}(X)\Big)
      }
      $}
    }{
    \raisebox{-4mm}{$
      \displaystyle
      1+
      \frac{\lambda^{ }_\theta
      \big( K;T(X),\vec u(X) \big)
      }{3}
      $}
    }
    \; .
\end{equation}
Having obtained $\lambda^{ }_i(K)$ by averaging over 
the system's space-time history, 
the invariant mass ($M$) and the transverse momentum ($\pT$) spectra of 
$\lambda^{ }_i$ are obtained in an analogous way. 
Namely, for our convention 
$
  K^\mu 
  = 
  \left(
    \sqrt{M^2 + \pT^2}\cosh y, 
    \pT \cos\phi, 
    \pT \sin \phi, 
    \sqrt{M^2 + \pT^2}\sinh y
  \right)
$, 
$\d^4 K = M\d M\,p_{\rm _T} \d p_{\rm _T}\d y \d \phi$, 
so $\lambda^{ }_i(M) $ and $\lambda^{ }_i(\pT) $ are given by, respectively,
\begin{eqnarray}
  \label{eq:lambda_i_M}
  \lambda^{ }_i(M) 
  & = &
  \frac{
  \raisebox{3mm}{$
    \displaystyle
    \int p_{\rm _T} \d p_{\rm _T} \d \phi \d y
    \ \lambda^{ }_i(K) \;
    {\cal N}(K)
    $}
  }{
  \raisebox{-3mm}{$
    \displaystyle
    \int  p_{\rm _T} \d p_{\rm _T} \d \phi \d y\  
    {\cal N}(K)
    $}
  }
  \; , \\
  \label{eq:lambda_i_pT}
  \lambda^{ }_i(\pT) 
  & = &
  \frac{
  \raisebox{3mm}{$
    \displaystyle
    \int M \d M \d \phi \d y
    \ \lambda^{ }_i(K) \;
    {\cal N}(K)
    $}
  }{
  \raisebox{-3mm}{$
    \displaystyle
    \int  M \d M \d \phi \d y \d \phi \d y\  
    {\cal N}(K)
    $}
  }
  \; .
  \\[-4mm] 
  \nonumber
\end{eqnarray}
\par
Since multiple Lorentz boosts are involved when calculating 
the averaged polarization coefficients for all fluid cells, 
it is natural to ask whether the rotational invariance of 
$\tilde \lambda$ given in 
Eq.~\eqref{eq:invlambda} can still hold after averaging 
over the entire fireball. 
The answer is yes. 
Let's consider a given lab-frame virtual photon 4-momentum $K$ 
and introduce the local quantities 
$\lambda_\theta(X)$, $\lambda_\phi(X)$ and ${\cal N}(X)$ 
as the polarization coefficients and the weight function 
for the fluid cell at space-time position $X\,$. 
As $\tilde \lambda$ is not directly defined in 
Eq.~\eqref{eq:defpol}, instead of using Eq.~\eqref{eq:avg_lambda}, 
the space-time averaged version of 
$\tilde\lambda$ should be obtained by 
its definition Eq.~\eqref{eq:invlambda}, but with the 
$\lambda_\theta$ and $\lambda_\phi$ being replaced by their 
respective values after averaging\footnote{%
  Practically and importantly, 
  $\lambda_\theta^{\rm av}$ and $\lambda_\phi^{\rm av}$ are 
  what an experiment would extract from the dilepton angular distribution 
  data $\d N/\d \cos\theta_{\ell}$. 
  Substituting those average values 
  in the original definition of $\tilde\lambda$, 
  yields our definition 
  of ${\tilde \lambda}^{\rm av}$ given by the first line 
  of Eq.~\eqref{eq:avg_inv}.
},  
i.e. 
(to lighten the notation and focus on the $X$ coordinate, we now omit 
implicit dependencies in the functions as 
given via Eqs.~\eqref{eq:avg_lambda} and \eqref{eq:defN})
\begin{eqnarray}
  \tilde \lambda^{\rm av} 
  & \equiv &
  \frac{
    \lambda^{\rm av}_\theta + 3\lambda^{\rm av}_\phi
  }{
    1-\lambda^{\rm av}_\phi
  } 
  \nonumber\\[2mm]
  & = &
  \frac{
  \raisebox{3mm}{$
    \displaystyle
    \int \d^4 X\,
    \big[\lambda^{ }_\theta(X)+3\lambda^{ }_\phi(X)\big] \,
    {\cal N}(X) 
  $}
  }{
  \raisebox{-3mm}{$
    \displaystyle
    \int \d^4 X\,
    \big[1-\lambda^{ }_\phi(X)\big] \, 
    {\cal N}(X)
  $}
  }
  \; .
  \label{eq:avg_inv}
\end{eqnarray}
The $M$ and $\pT$ spectra for $\tilde \lambda$ 
are obtained the the same way as in 
Eqs.~\eqref{eq:lambda_i_M} and 
\eqref{eq:lambda_i_pT}. 
The weight function ${\cal N}(X)$ can be worked out directly 
from Eq.~\eqref{eq:defN} and Eq.~\eqref{eq:d4rate} in terms of 
$\rho^{ }_{\rm V}(X)$ and $\rho^{ }_\Delta(X)$, 
which is found to be proportional to the common denominator $D$ of 
the polarization coefficients listed in 
Eqs.~\eqref{eq:ltheta}--\eqref{eq:lperpthph}. 
Consequently, several cancellations occur 
in Eq.~\eqref{eq:avg_inv} 
leading to an explicit representation 
which is reminiscent of 
the unintegrated version from  Eq.~\eqref{eq:invlambda}, 
namely 
\begin{widetext}
\begin{equation}   
  \tilde \lambda^{\rm av}
  \; = \;
  \frac{
  \raisebox{3mm}{$
    \displaystyle
    3(1- 4\xi)
    \int \d^4 X\,
    \nB(X)\left[ 
      \chi^{ }_z(X)+\chi^{ }_x(X) - \chi^{ }_y(X)-\frac{1}{3} 
    \right] \rho^{ }_\Delta(X)
  $}
  }{
  \raisebox{-3mm}{$
    \displaystyle
    \int \d^4 X\,
    \nB(X)\left[\,
      \frac{4}{3}(1+2\xi)\,\rho^{ }_{\rm V}(X) - 
      (1-4\xi)\left( \chi^{ }_z(X)+\chi^{ }_x(X) - 
      \chi^{ }_y(X)-\frac{1}{3}\right) \rho^{ }_\Delta(X)
    \,\right]
  $}
  }
  \; .
\end{equation}
\end{widetext}

Recall that $\vec u_*(X)$ denotes the local fluid velocity 
of different ``cells'' for a given 
virtual photon momentum $K\,$,  
which enters in the above 
only via the functions $\chi_i(X)$; 
the local spectral functions, $\rho^{ }_{\rm V}(X)$ and 
$\rho^{ }_\Delta(X)$, and Bose-Einstein 
distribution, $\nB(X)$, can be expressed in a 
covariant manner, cf. the discussion above Eq.~\eqref{eq:d8rate}. 
Both the numerator and denominator in Eq.~\eqref{eq:avg_inv} 
involve the particular combination 
$\chi_z(X) + \chi_x(X) - \chi_y(X)$, 
which remains invariant under a $y$-rotation 
between different $\gamma^*$ rest frames. 
The invariance is clearly preserved when summing over cells, 
as well as when integrating over the full $\pT$ range 
to obtain the $M$ distribution. 
This completes the proof of the frame-invariance 
(CS/HX) of ${\tilde \lambda}^{\rm av}$.

%
\section{Multi-stage hydrodynamic model}
\label{sect:hydro}

\subsection{iEBE-MUSIC}

As discussed above, we focus on dilepton polarization 
in heavy-ion collisions, 
which are highly dynamical 
and complex systems that are usually described 
using multi-stage hybrid models. 
In this work, we adopt the iEBE-MUSIC 
framework~\cite{Schenke:2010nt,Schenke:2010rr,Paquet:2015lta} 
to simulate the space-time evolution of 
Pb+Pb collisions with $\sqrt{s_{\rm NN}} = 5.02$~TeV at the LHC.

At high collisional energies, 
the initial state of a nuclear projectile has been described within 
the colour-glass condensate (CGC) 
framework~\cite{Iancu:2002xk}. 
In the present study, the primordial evolution from  
$\tau = 0^+$ to $0.1~\rm fm$, is simulated with the IP-Glasma 
model~\cite{Schenke:2012wb,Schenke:2012hg}. 
This model is based on the CGC effective field theory, 
where small-$x$ gluons are described by the classical action 
\begin{align}
  S_{\rm CGC} 
  \; \equiv \; 
  \int d^4 X 
  \Big( 
    -\frac{1}{4}F^a_{\mu\nu}F_a^{\mu\nu} 
    + J^{\mu}_a A_{\mu}^a
    \, \Big)
  \; .
\end{align}
Here $F^a_{\mu\nu}$ is the field-strength tensor with color index $a$, 
and $J^{a\mu}$ denotes the color current from the large-$x$ partons 
which can generate the soft gluons. 
Consequently, the evolution of the gluon field before the collision 
is governed by the classical Yang-Mills equations with source terms 
as follows:
\begin{eqnarray}
  \big[D_{\nu},F^{\mu\nu}\big] 
  & = & J^{\mu} \\
  & \equiv & 
  \delta^{\mu\pm} \rho_{A(B)}(x^{\mp},\textbf{x}_{\perp})
\end{eqnarray}
where $D_{\nu} =\partial_{\nu} - igA_{\nu}$ 
is the covariant derivative with the convention 
$A^{\mu}=A^{\mu}_{a}t^{a}$, and $t^{a}$ are the SU(3) generators. 
Denoting lightcone coordinates 
$x^{\pm} \equiv (t \pm z)/\sqrt{2}\,$,
$\rho_{A(B)}(x^{\mp},\textbf{x}_{\perp})$ represents 
the fluctuating color charge density of target and 
projectile nuclei, estimated from the IP-SAT 
model~\cite{Kowalski:2003hm}. 
For large nuclei, 
we first use a Woods–Saxon distribution to sample 
the spatial positions of nucleons in the target and 
projectile. 
Each nucleon is further considered as three hot spots 
to include sub-nucleonic fluctuations. 
After the initial configuration of nucleons 
and sub-nucleonic hot spots are fixed, 
the IP-SAT model is employed to determine 
the local saturation scale $Q^{ }_s$ in the transverse plane. 
Then, the colour charge distribution of the target 
and projectile nuclei is sampled according to 
the two-point correlation function of the 
colour charge density, i.e. 
\begin{align}
  \big\langle \rho_A^a(X)\rho_A^b(Y) \big\rangle 
  \; = \; 
  g^2 \mu_A^2(X,Y)\delta^{ab}\delta^2(X-Y)
  \; , 
\end{align}
where the gluon distribution function 
$g^2 \mu_A^2 \propto  Q^{ }_s\,$. 
After the collision, 
the gluon fields continue to evolve according to the 
sourceless Yang–Mills equations until 
the onset of the pre-equilibrium stage. 

At this point, the energy-momentum tensor 
$T^{\mu\nu}_{\rm IP}$ is constructed from 
the chromoelectric and chromomagnetic fields 
in the IP-Glasma model. 
For more details of IP-Glasma, 
refer to Refs.~\cite{Schenke:2012wb,Schenke:2012hg}. 
The $T^{\mu\nu}_{\rm IP}$ is also used as the 
input to the pre-equilibrium stage. 
The dynamical evolution in the pre-equilibrium stage, 
is simulated with the \kompost 
model~\cite{Kurkela:2018wud,Kurkela:2018vqr}, 
which is active for a proper-time window between $0.1$ and $0.8$~fm. 
The lower limit is consistent with a saturation momentum 
$Q^{ }_s \sim$~2~GeV, 
and the upper limit is consistent with that used in recent studies 
such as Refs.~\cite{Kurkela:2018vqr,Gale:2021emg}. 
In \kompostend, the full energy–momentum tensor 
$T_{K}^{\mu\nu}(\tau,\mathbf{x})$ 
is decomposed into a locally homogeneous background  
$\bar{T}_{K}^{\mu\nu}(\tau,\mathbf{x})$ 
and perturbations 
$\delta T_{K}^{\mu\nu}(\tau,\mathbf{x})$:
\begin{align}
  T_{K}^{\mu\nu}(\tau,\mathbf{x}) 
  \; = \; 
  \bar{T}_{K}^{\mu\nu}(\tau,\mathbf{x}) 
  + 
  \delta T_{K}^{\mu\nu}(\tau,\mathbf{x}).
\end{align}
The local homogeneous background 
$\bar{T}_{K}^{\mu\nu}(\tau,\mathbf{x})$ follows 
a universal curve obtained from studies of 
the hydrodynamic attractor, 
while the perturbation 
$\delta T_{K}^{\mu\nu}(\tau,\mathbf{x})$ is computed 
from linear response functions estimated within 
pure-gluonic QCD effective kinetic theory. 
Both the background and the perturbation depend 
on the shear viscosity to entropy density $\eta/s$, 
which is set to a constant value of $0.12$ 
here and in the subsequent hydrodynamic evolution.

At $\tau = 0.8~{\rm fm}$, 
the \kompost energy-momentum tensor 
$T_{K}^{\mu\nu}(\tau,\mathbf{x})$ 
is matched to relativistic viscous hydrodynamics, 
which is simulated by the MUSIC 
code~\cite{Schenke:2010nt,Schenke:2010rr,Paquet:2015lta}. 
We adopt the Landau matching procedure, 
$u_\mu T_{K}^{\mu \nu} \equiv \varepsilon u^{\nu}$, 
to determine the initial energy density $\varepsilon $ 
and flow velocity $u^{\mu}$. 
Due to the conformal assumption in the \kompost model, 
and in order to ensure that $T^{\mu\nu}(\tau,\mathbf{x})$ 
is exactly the same on both sides of the switching 
hypersurface between the pre-equilibrium and 
hydrodynamic stages, the initial shear-stress tensor 
$\pi^{\mu\nu}$ and bulk viscous pressure $\Pi$ 
are initialized via
\begin{eqnarray}
  \pi^{\mu\nu} 
  & = &
  T_{K}^{\mu\nu}(\tau,\mathbf{x})
  -
  \frac{4}{3}\varepsilon \, u^{\mu} u^{\nu} 
  +
  \frac{\varepsilon }{3}g^{\mu\nu} \\[2mm]
  \Pi 
  & = &
  \frac{\varepsilon }{3} - p(\varepsilon )
  \; ,
\end{eqnarray}
where $p(\varepsilon )$ denotes the pressure determined from the 
lattice equation of state (EOS) determined by the 
HotQCD Collaboration~\cite{HotQCD:2014kol}. 
After the Landau matching, 
the system is evolved according to energy–momentum 
conservation and the Denicol–Niemi–Moln\'{a}r–Rischke 
(DNMR) 
second-order viscous hydrodynamic 
formalism~\cite{Denicol:2012cn}. 
In this study, 
we use a temperature-dependent bulk viscosity 
to entropy density ratio defined by 
\begin{align}
  \frac{\zeta}{s}(T) 
  \; = \; 
  \begin{cases}
  B_{\text{norm}}
  \exp\!\left[-\dfrac{(T-T_{\text{peak}})^2}{B_1^{\,2}}\right], 
  & T < T_{\text{peak}}, \\[4mm]
  B_{\text{norm}}
  \exp\!\left[-\dfrac{(T-T_{\text{peak}})^2}{B_2^{\,2}}\right], 
  & T > T_{\text{peak}} \, ,
  \end{cases}
\end{align}
where 
$B_{\text{norm}}=0.175$, 
$B_1 = 0.01~\rm GeV$, 
$B_2 = 0.12~\rm GeV$ and 
$T_{\text{peak}}=0.16~\rm GeV$.

Once the local energy density in a fluid cell drops to 
$\varepsilon =0.18~{\rm GeV/fm^3}$, 
the Cooper-Frye prescription with 14-moment 
viscous corrections~\cite{Ryu:2015vwa, Schenke:2020mbo} 
is used to convert the fluid into thermal hadrons. 
These hadrons are then fed into the UrQMD model~\cite{Shen:2014vra} 
to undergo further hadronic scatterings and decays.

\subsection{Chemical equilibrium}

The initial condition of the multi-stage model 
approximates the QGP as a purely gluonic system, 
which has been very successful in reproducing the final hadronic 
multiplicities and collective flow phenomena observed in HICs. 
However, the presence of quarks and antiquarks is required to calculate 
the dilepton production during the pre-equilibrium stage, 
where they are gradually produced 
via gluon splitting until chemical equilibrium is 
established after a finite time. 
Moreover, in the current setup, the DPR is calculated 
at NLO assuming local thermal and chemical equilibrium. 
A first-principle description of pre-equilibrium 
stage dileptons is not readily available. 
In this paper, following our previous work~\cite{Wu:2024pba}, 
the pre-equilibrium DPR is estimated in 
a phenomenological way: 
(i) 
obtaining an effective temperature $T_\text{pre-eq}$ from 
the local energy density by Landau matching, 
using the same EOS as in the subsequent hydrodynamical stage; 
(ii) 
introducing a phenomenological suppression factor 
$\sf(T_\text{pre-eq},\tau)$ to account for the 
limited dilepton production caused by 
the reduced quark abundance during 
the pre-equilibrium stage, where $\tau$ denotes 
the proper time 
and $T_\text{pre-eq}$ is an effective temperature; and 
(iii) 
using the equilibrium DPR with temperature being 
$T_\text{pre-eq}$ 
and considering possible implementations of 
the $\sf(T_\text{pre-eq},\tau)$ modification factor.

Specifically, quark and antiquark distribution function 
in the pre-equilibrium stage are assumed\footnote{\label{foot:distortions}%
  We neglect momentum dependence in the suppression factor, 
  and are thus essentially neglecting any 
  non-trivial pre-equilibrium distortion 
  to the shape of the distribution function.
}
to be given by 
$
  f_{q/\bar q}^\text{pre-eq} 
  = 
  \sf(T_\text{pre-eq}\, , \tau ) 
  \cdot 
  f_{q/\bar q}^{\rm eq} \,
$. 
Because LO dilepton production uniquely contains 
the annihilation channel $q\bar q\to \gamma^*$,
we anticipate the LO DPR is corrected by 
the quark suppression as 
\begin{equation}
  {\rm d}R^\text{pre-eq}_{\rm LO} 
  \; = \; 
  \sf^2(T_\text{pre-eq},\tau) 
  \cdot
  {\rm d}R^{\rm eq}_{\rm LO}
  \, .
\end{equation}
\par
The suppression factor acts differently on the NLO channels. 
For example, Compton scattering $gq\to \gamma^* q$ involves 
only one quark in its initial state
so the rate of this process $R\sim \sigma(gq\to \gamma^* q) f_q f_g$ 
should be less ``corrected'' by $\sf(T_\text{pre-eq}\,,\tau)$ than 
that of $q\bar q$ annihilation\footnote{%
  For $gq\to \gamma^* q$, the final-state quark also contributes 
  a Pauli blocking factor $1-f_q$ to the rate, but 
  we neglect   intricacies of this type in the suppression factor. 
}.
Another NLO process, namely modified annihilation 
$q\bar q \to \gamma^* g$, 
receives approximately the same modification due to quark suppression 
as the LO process $q\bar q \to \gamma^*$. 
Although it is theoretically not well-defined to distinguish each channel 
in the thermal spectral function $\rho^{\mu\nu}$ (individual channels 
may contain kinematic divergences which only cancel when combined), 
since all NLO processes involve either one or two initial (anti)quarks, 
we expect the overall NLO DPR to be suppressed less than 
the LO DPR, characterized by a phenomenological parameter~$\alpha\,$. 
Having this in mind, we adopt 
the following differential DPR for 
the pre-equilibrium stage
\begin{eqnarray}
  \d R^\text{pre-eq} 
  & = &
  {\rm SF}^2(T_\text{pre-eq}\,,\tau)
  \cdot \d R_{\rm LO}^{\rm eq}
  \nonumber\\[2mm]
  & + &  
  {\rm SF}^\alpha(T_\text{pre-eq}\,,\tau)
  \cdot \d R_{\rm NLO}^{\rm eq}
  \; ,
  \label{eq:dr_pre-eq}
\end{eqnarray}
with the suppression index $ \alpha \in (1,2)$.

%
\section{Results}\label{sect:results}

\subsection{Polarization coefficients in different $\gamma^*$\\[1mm] rest frames}

\begin{figure}[t]
\centering
\includegraphics[width=1.0\linewidth]{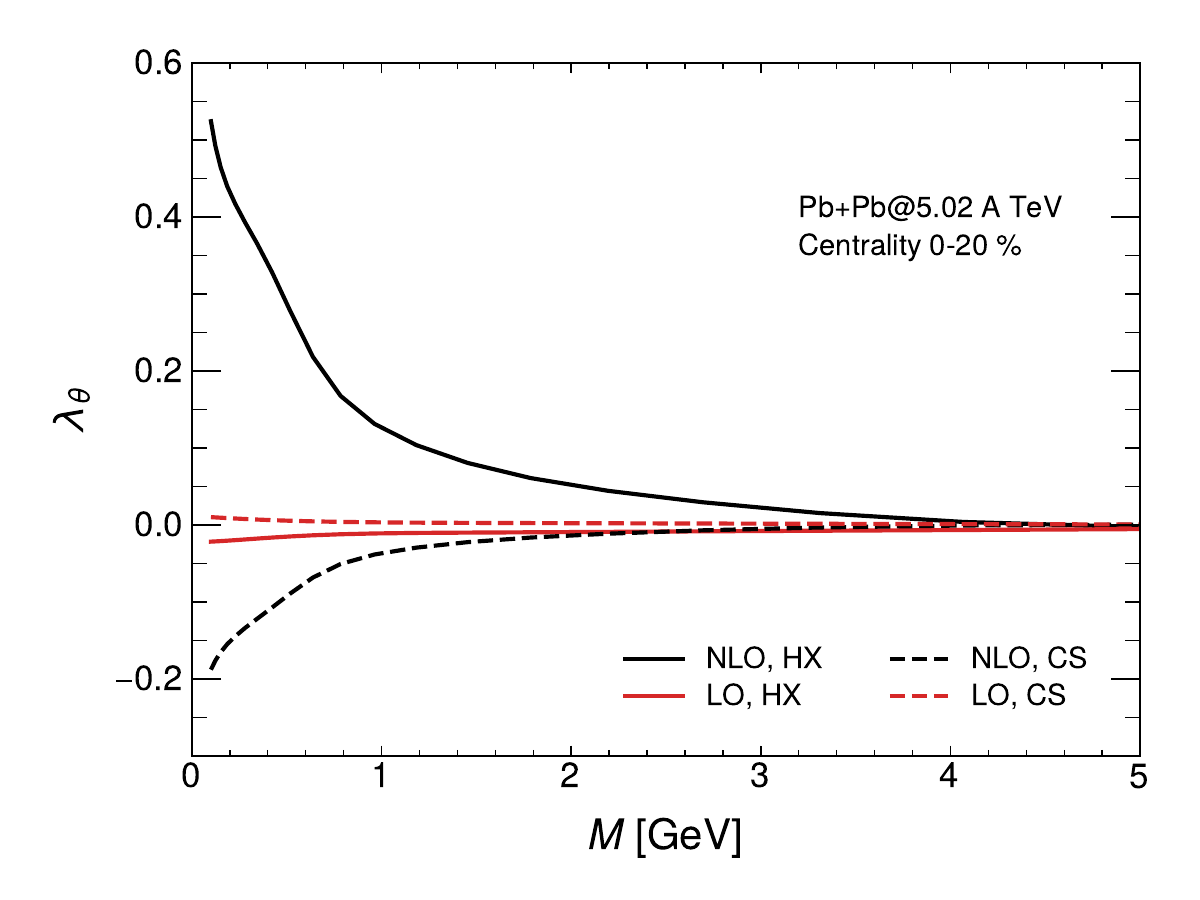}
\vspace{-6mm}
\caption{
  Thermal LO (red curves) and NLO (black curves) prediction of the 
  dilepton $\lambda_\theta$ invariant mass spectrum 
  (integrated over the entire $p_T$ range using eq.~\eqref{eq:lambda_i_M}) 
  in the HX 
  (solid curves) and CS (dashed curves) frames. 
  The role of NLO processes manifest themselves visibly in the HX frame.
}   
\label{fig:cs_nlo}
\end{figure}

Figure~\ref{fig:cs_nlo} shows our model prediction for  
$\lambda_\theta$ as a function of 
the dilepton invariant mass $M$ in both HX and CS frames, 
using the LO and NLO spectral functions 
(in this work, a fixed $\alpha_s = 0.3$ is used for the NLO results, 
and all values are evaluated at mid-rapidity ($y =0$)). 
As in our previous work~\cite{Wu:2024vyc}, 
a significant difference of the HX-frame $\lambda_\theta(M)$ 
is seen when the NLO corrections are included.\footnote{%
  It should be noted that at LO, where explicit expressions 
  for $\rho_{_{\rm T,L}}$ are available 
  (see e.g.~Re.~\cite{Churchill:2023vpt}), 
  both spectral functions vanish in the limit $M\to 0$ 
  for kinematic reasons. However, the ratio 
  determining $\lambda^{\rm HX}_\theta$ in 
  eq.~\eqref{eq:lth_approx} is actually finite 
  in this limit, and given by
  \begin{eqnarray}\label{eq:F(x)}
      \lim_{M\to 0}
      \frac{
       \rho_{_{\rm T}}  - \rho_{_{\rm L}}
      }{
       \rho_{_{\rm T}}  + \rho_{_{\rm L}}
      } \bigg|_{\rm LO}
      & \equiv & 
      F(k/T) \; , 
  \end{eqnarray}
  where $F(k/T)$ is a function which can be 
  given in terms of polylogarithms. 
  In particular, $F(x) < 0$ and behaves as:
  \begin{align}
  F(x) =
  \begin{cases}
  - 3 \log 2/(2x) & \text{for} \ \  x \gg 1 \\[4pt]
  - x^2/160 & \text{for} \ \  x \ll 1 
  \end{cases}
  \; .
  \end{align}
  And $F_{\rm min} \approx - 0.06$. 
  This explains the LO curves in Fig.~\ref{fig:cs_nlo}, 
  and similar plots in the literature which rely on processes such 
  as $q \bar{q} \to \gamma^*$ or $\pi^+ \pi^- \to \gamma^*$. 
  However, Eq.~\eqref{eq:F(x)} is an artifact of the 
  vanishing phase space for such processes when $M$ is small 
  and both spectral functions vanish. 
}
In CS frame, this difference is still present, 
but is significantly smaller.  
It is also observed that both LO and NLO version of 
$\lambda_\theta(M)$ 
change their sign when switching from HX to CS frame. 

%
%

\begin{figure}[t]
\centering
\includegraphics[width=1.0\linewidth]{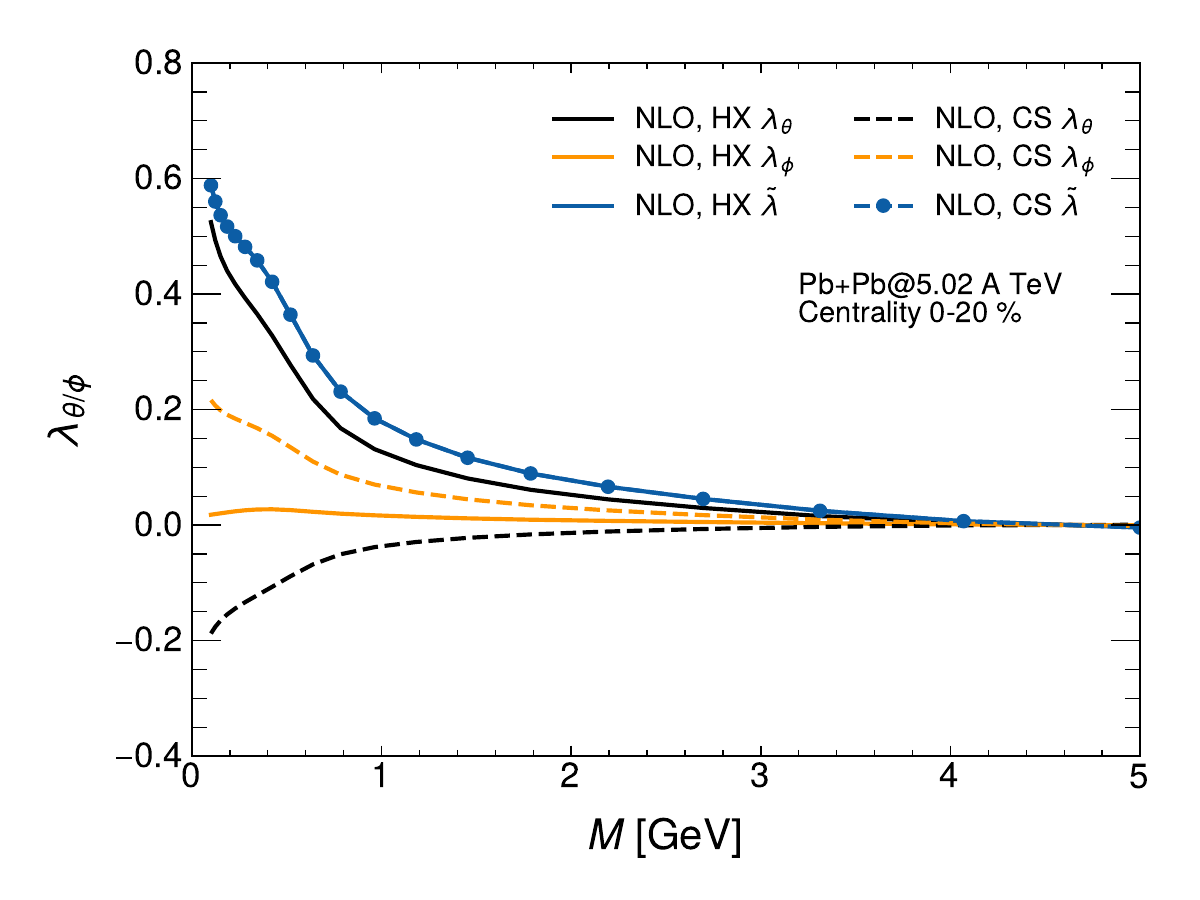}
\vspace{-6mm}
\caption{
  Thermal NLO predictions for polarization coefficients 
  $\lambda_\theta(M)$ (black curves) 
  and $\lambda_\phi(M)$~(orange curves) in 
  the HX (solid) and CS (dashed) frames, 
  and the invariant combination $\tilde \lambda(M)$ calculated 
  in both frames. 
  The overlapping of $\tilde\lambda^{\rm HX}(M)$ and 
  $\tilde\lambda^{\rm CS}(M)$ curves demonstrates that $\tilde\lambda$ 
  is continues to be invariant even after 
  integrating over the fireball's spacetime history
  and the dilepton's $\pT$~(see Eq.~\eqref{eq:avg_inv}). 
  }
\label{fig:lambda_NLO}
\end{figure}

The NLO prediction of another polarization coefficient, 
namely $\lambda_\phi$ as a function of $M$, 
is shown in Fig.~\ref{fig:lambda_NLO}, 
along with $\lambda_\theta$ and $\widetilde \lambda$, 
in both HX and CS frame. 
The near-zero $\lambda_\phi$ in HX frame increases sizably when it is 
calculated in the CS frame. 
Despite all the differences observed for 
$\lambda_\theta$ and $\lambda_\phi$ between these two frames, 
$\widetilde \lambda$ remains exactly invariant 
as promised by Eq.~\eqref{eq:invlambda} even after averaging 
over the entire evolution history of the fireball 
as discussed in Sec.~\ref{sect:superp_coeff}.

%
%

\par
\begin{figure}[t]
\centering
\includegraphics[width=1.0\linewidth]{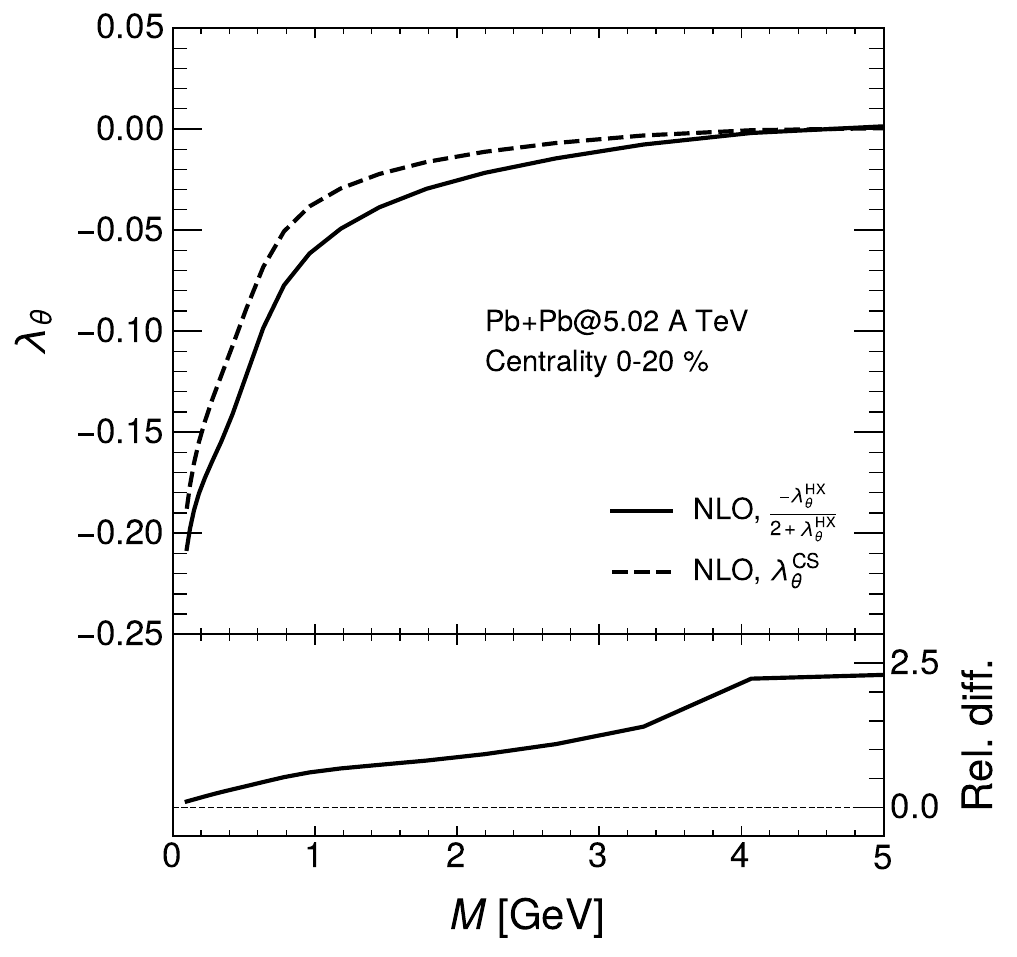}
\vspace{-6mm}
\caption{
  Upper panel: comparison between $\lambda_\theta^{\rm CS}(M)$~(solid line)
  and the right side of Eq.~\eqref{eq:lth_hx_cs}~(dashed line), 
  both calculated from hydrodynamics using NLO thermal rates. 
  Lower panel: the relative difference between the two curves shown 
  in the upper panel, with 
  $
    \text{Rel. diff.} \equiv 
      \lbk \lambda_\theta^{\rm CS} - \frac{-\lambda_\theta^{\rm HX}}{2+\lambda_\theta^{\rm HX}}\rbk\Big/\frac{-\lambda_\theta^{\rm HX}}{2+\lambda_\theta^{\rm HX}}
  $. 
  This difference is a result of 
  the dynamical nature of the system; see the main text for details.
}
\label{fig:diff_frames}
\end{figure}

To investigate the difference of polarization coefficients in 
these two frames, let's consider again a uniform background with 
$\vec u = 0$ so $\vec u_* = -\vec k/M\,$. For mid-rapidity dileptons, 
$\vec k\cdot \hat n =0$ so Eq.~\eqref{eq:ezcs} reads 
$\vec e_z^{\rm CS} = \hat n \,$. 
In the $m_\ell = 0$ limit, Eq.~\eqref{eq:ltheta} yields
\begin{equation}\label{eq:lth_hx_cs}
  \lambda_\theta^{\rm CS} 
  {
  \Big|_{\vec{u}=0}
  }
  \; = \; 
  \frac{
    \rho_{_{\rm T}} - \rho_{_{\rm L}}
  }{
    \rho_{_{\rm L}} + 3\rho_{_{\rm T}}
  } 
  \; = \; 
  -\, \frac{
    \lambda_\theta^{\rm HX}
  }{
    2+\lambda_\theta^{\rm HX}
  }{
  \Bigg|_{\vec{u}=0}
  }
  \;.
\end{equation}
Although Eq.~\eqref{eq:lth_hx_cs} was derived for $\vec{u}\to 0\,$, 
it appears to hold approximately for realistic predictions as 
can be seen by comparing the corresponding curves in 
Fig.~\ref{fig:lambda_NLO} with one another.
Observing from Fig.~\ref{fig:lambda_NLO}, 
this equation indeed explains the different behaviours of 
$\lambda_\theta$ 
(and thus $\lambda_\phi$, using Eq.~\eqref{eq:invlambda}) 
in the HX and CS frames: 
a flipped sign and a smaller magnitude in the CS frame. 
Nevertheless, since the QGP is generally not static, 
our prediction indeed deviates from Eq.~\eqref{eq:lth_hx_cs}. 
This deviation is shown in Fig.~\ref{fig:diff_frames}, 
where 
$
  \lambda_\theta^{\rm CS}(M) 
  > 
  -\frac{\lambda_\theta^{\rm HX}(M)}{2+\lambda_\theta^{\rm HX}(M)}
$ 
holds for all values of $M$. 
\par
A quantitative explanation of this 
inequality requires knowing the analytical form of 
the spectral functions, 
which is not available for the NLO result\footnote{%
  Note that we do not include the relative 
  order $\sqrt{\alpha_s}$ corrections to the 
  ``soft dilepton'' 
  rate~\cite{Ghiglieri:2013gia,Ghiglieri:2014kma}. 
} 
(the latter must be computed numerically, 
and is available in tabulated 
form~\cite{dileptoncode,zenodo_link_GJ}). 
Qualitatively, the primary difference between 
a real QGP and an ideal QCD plasma in a static box 
is the QGP's longitudinal expansion. 
Let's consider the correction to Eq.~\eqref{eq:lth_hx_cs} brought 
by a longitudinal flow velocity $u^\mu = (\sqrt{1+u_l^2},0,0,u_l)$. 
Performing the analysis again, we have, for mid-rapidity dileptons
\begin{equation}\label{eq:lth_hx_ul}
  \lambda_\theta^{\rm HX} 
  \; = \; 
  \frac{
    \rho_{_{\rm T}}-\rho_{_{\rm L}}
  }{
    \rho_{_{\rm T}} + \rho_{_{\rm L}}
  } 
  -
  \frac{M^2}{k^2} 
  \frac{
    (2\rho_{_{\rm T}} + \rho_{_{\rm L}})(\rho_{_{\rm T}} - \rho_{_{\rm L}})
  }{
    (\rho_{_{\rm T}} + \rho_{_{\rm L}})^2
  }
  u_l^2 + O(u_l^4) 
  \, ,
\end{equation}
and
\begin{equation}\label{eq:lth_cs_ul}
  \lambda_\theta^{\rm CS} 
  \; = \; 
  -
  \frac{
    \rho_{_{\rm T}}-\rho_{_{\rm L}}
  }{
    3\rho_{_{\rm T}} + \rho_{_{\rm L}}
  } 
  +
  \frac{4M^2}{k^2}
  \frac{
    (2\rho_{_{\rm T}} + \rho_{_{\rm L}})(\rho_{_{\rm T}} - \rho_{_{\rm L}})
  }{
    (3\rho_{_{\rm T}} + \rho_{_{\rm L}})^2
  }
  u_l^2 + O(u_l^4)
  \, .
\end{equation}
From Eq.~\eqref{eq:lth_hx_ul} and Eq.~\eqref{eq:lth_cs_ul}, 
we can write down the $u_l$ correction of Eq.~\eqref{eq:lth_hx_cs} as
\begin{equation}
  \lambda_\theta^{\rm CS} 
  \; = \; 
  - \frac{\lambda_\theta^{\rm HX}}{2+\lambda_\theta^{\rm HX}} 
  + \frac{M^2}{k^2}
  \frac{
    \lambda_\theta^{\rm HX}(3+\lambda_\theta^{\rm HX})
  }{
    (2+\lambda_\theta^{\rm HX})^2
  }u_l^2 
  + O(u_l^4)
  \; .
\end{equation}
In practice, the lowest-order correction always has 
the same sign as $\lambda_\theta^{\rm HX}$ (which is positive), 
regardless of the sign of $u_l$. 
This confirms our observation that 
$
  \lambda_\theta^{\rm CS}(M) 
  >
  -\frac{\lambda_\theta^{\rm HX}(M)}{2+\lambda_\theta^{\rm HX}(M)}
$. 
Although this analysis is simplistic, 
neglecting the additional transverse flow and that $u_l$ can be $O(1)$ 
(for large space-time rapidities), 
it suggests that the quantity 
$
  \lambda_\theta^{\rm CS}(M)
  +
  \frac{\lambda_\theta^{\rm HX}(M)}{2+\lambda_\theta^{\rm HX}(M)}
$ 
encodes the longitudinal flow strength of the QGP. 

%
%

\par
Equations~\eqref{eq:lth_hx_ul} and \eqref{eq:lth_cs_ul} 
also suggest that $\lambda_\theta^{\rm CS}$ is more sensitive 
to flow conditions than its HX frame counterpart. 
Meanwhile, if only the intrinsic properties of the plasma are considered 
(that is, only the microscopic dilepton production channels but 
without any flow velocity $\vec u$), 
$\lambda_\theta^{\rm HX}$ has a larger magnitude as 
shown in Fig.~\ref{fig:diff_frames}. 
In practice, these observations suggest using 
$\lambda_\theta^{\rm CS}$ to probe anisotropic elements of the plasma
such as the plasma anisotropy and  external EM field, 
and turning to  $\lambda_\theta^{\rm HX}$, 
to quantify QGP intrinsic properties, including temperature.

\subsection{Pre-equilibrium dilepton polarization}\label{ssect:pre-eq}

%
\begin{figure}[t]
\centering
\includegraphics[width=1.0\linewidth]{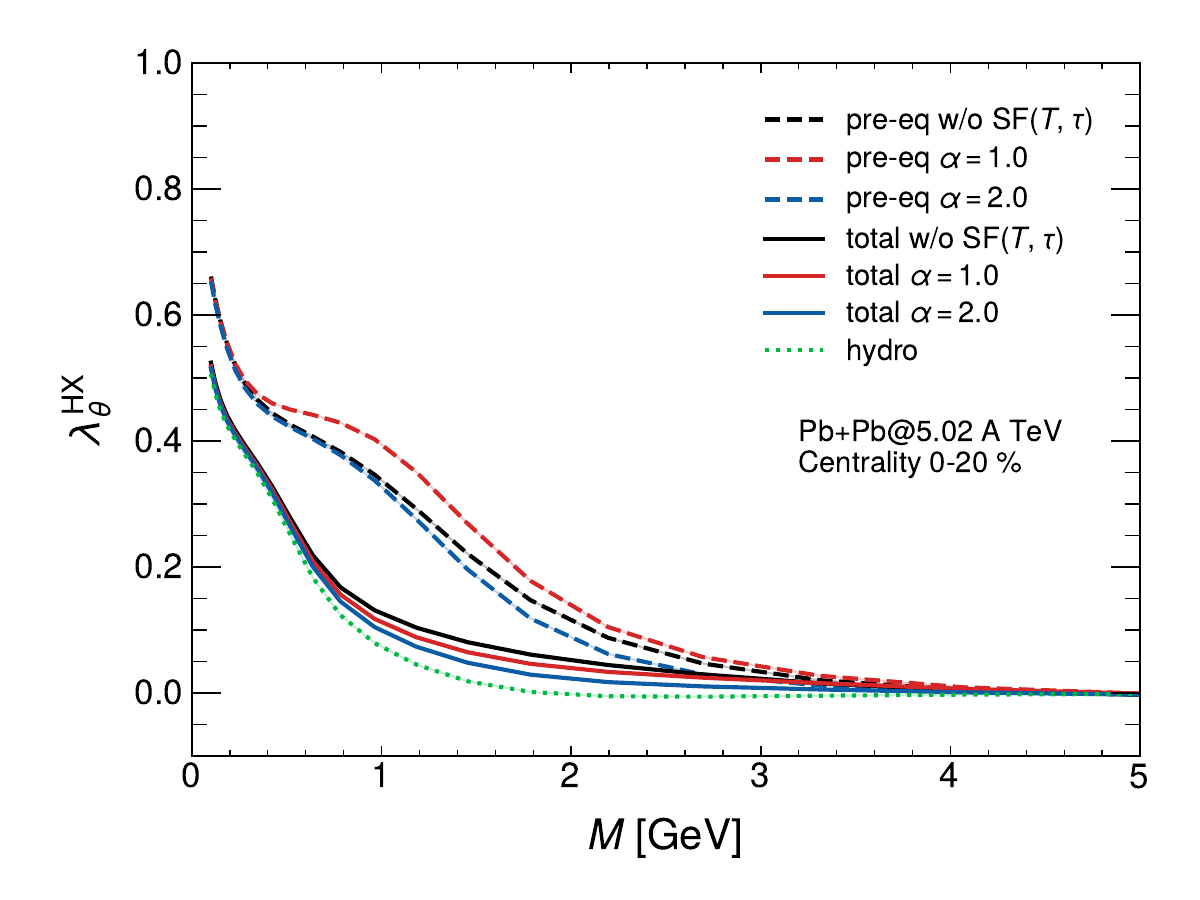}
\vspace{-6mm}
\caption{
  The polarization coefficient 
  $\lambda^{\rm HX}_\theta(M)$ of the pre-equilibrium dileptons 
  (dashed lines) 
  and the total dileptons (solid lines), 
  for suppression index $\alpha = 1,2$ and for no pre-equilibrium 
  quark suppression is used (indicated by ``w/o $\sf(T,\tau)$"). 
  Here, ``total dileptons" refer to pre-equilibrium dileptons mixed 
  with dileptons produced in the hydrodynamic stage (dotted green line).  
}
\label{fig:lth_M_sf}
\end{figure}
%

Figure~\ref{fig:lth_M_sf} shows the polarization coefficient 
for three different scenarios of the pre-equilibrium stage 
using Eq.~\eqref{eq:dr_pre-eq} and studying a 
suppression index $\alpha = 1$, $2$ (for the NLO rate), 
and using no suppression factor (for neither LO nor NLO rates). 
We plot $\lambda_\theta^{\rm HX}(M)$
for both the pre-equilibrium value alone 
and the total value obtained after averaging 
with the hydrodynamic stage result.
The qualitative behaviour of $\lambda_\theta^{\rm HX}(M)\,$, 
due to the pre-equilibrium component, 
is found to be robust against 
variations of the suppression index $\alpha$. 
\par
\begin{figure}[t]
\centering
\includegraphics[width=1.0\linewidth]{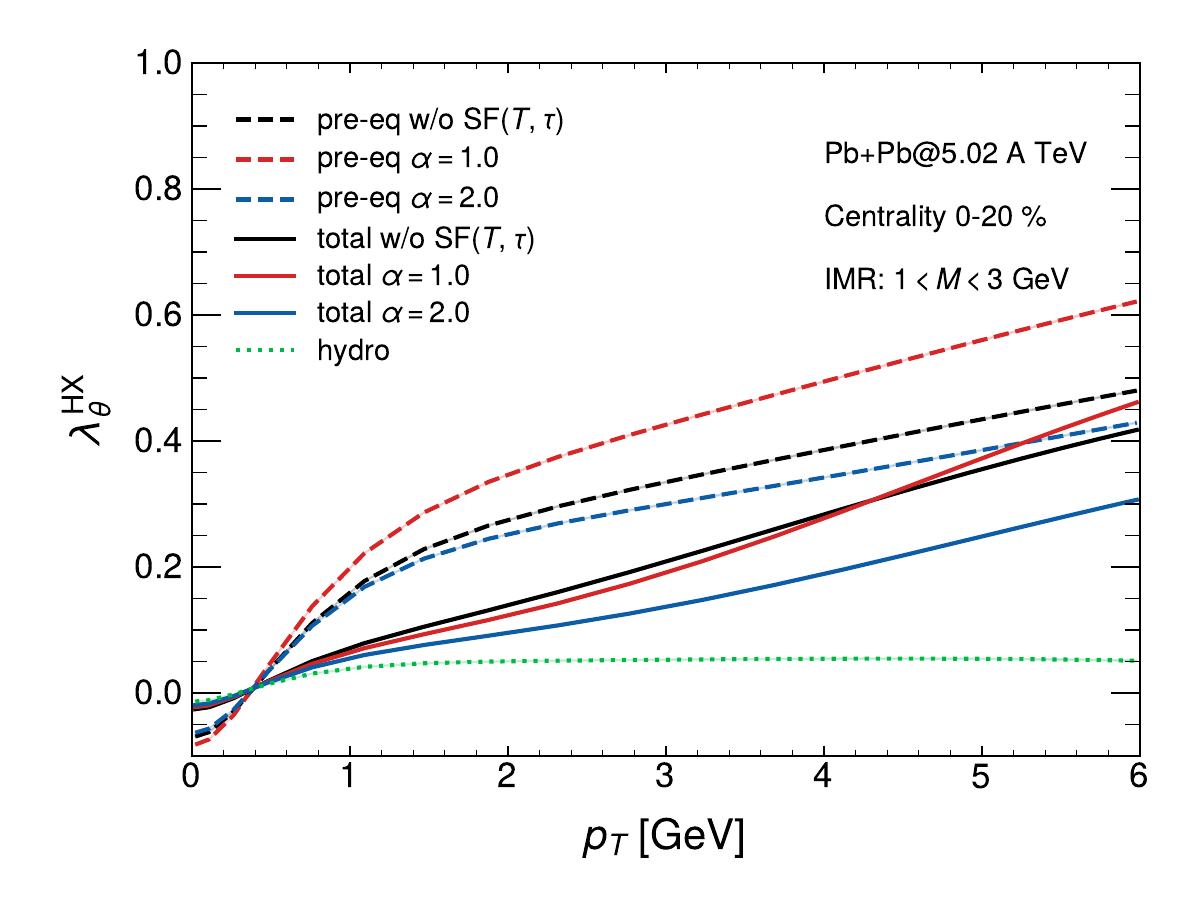}
\vspace{-6mm}
\caption{
  This plots shows the $\lambda^{\rm HX}_\theta(\pT)$ 
  coefficient of the pre-equilibrium dileptons (dashed lines), 
  of the total dileptons (solid lines), and of 
  the hydrodynamic-stage dileptons (dotted green line) in the IMR. 
  The linestyles and the colours are encoded in the same way 
  as in Fig.~\ref{fig:lth_sf_M}. 
}
\label{fig:lth_pT_IMR}
\end{figure}

Nevertheless, the exact value of the total dilepton polarization 
coefficient $\lambda_\theta^{\rm HX}$ in the IMR is appreciably influenced 
by dileptons produced in the pre-equilibrium stage. 
This observation suggests the feasibility of 
extracting pre-equilibrium properties of the 
fireball by studying the 
$\lambda_\theta^{\rm HX}(\pT)$ 
spectrum in the $1 <M<3~{\rm GeV}$ window. 
This is demonstrated in Fig.~\ref{fig:lth_pT_IMR} by varying 
the NLO suppression index $\alpha$,
and by comparing with the scenario
where no suppression factor is applied.\footnote{%
  Note that in Fig.~\ref{fig:lth_pT_IMR}, 
  the dotted green line for hydrodynamic dileptons 
  and the two black lines 
  for the case ``without $\sf(T,\tau)$'' 
  are calculated using the same settings as in 
  the lower panel of the Fig.~3 
  in Ref.~\cite{Wu:2024vyc}, and are therefore identical.
} 
The sensitivity to such effects is enhanced 
with increasing $\pT\,$.

%
%

\par 
To better understand the behaviour of 
the polarization coefficient in different settings, 
it is beneficial to keep in mind 
the two following effects, also discussed in Ref.~\cite{Wu:2024vyc}:
\begin{enumerate}
    \item {\it Gluon effect}: 
    When comparing LO and NLO results for dileptons, 
    bear in mind that the emission of a gluon can flip 
    the helicity\footnote{%
      As a consequence of angular momentum conservation, at LO the total 
      spin of the $q \bar q$-pair must match the spin of the $\gamma^*$. 
      If the quark and antiquark have opposite helicities (e.g. 
      $q_{_{\rm L}} {\bar q}_{_{\rm R}}$ or 
      $q_{_{\rm R}} {\bar q}_{_{\rm L}}$) the $\gamma^*$ is produced 
      in a transverse polarization state (helicity $\pm 1$). However 
      if the quark and antiquark have the same helicity (e.g. 
      $q_{_{\rm L}} {\bar q}_{_{\rm L}}$ or 
      $q_{_{\rm R}} {\bar q}_{_{\rm R}}$) the $\gamma^*$ is produced in 
      a longitudinal polarization state (helicity $0$). 
      However, at NLO the additional gluon can carry away spin 
      allowing to flip the helicity of the quark or antiquark. 
      In particular, when the gluon is emitted collinearly, the process 
      becomes kinematically similar to the LO process but with an 
      additional soft gluon.
    } 
    of the quarks, 
    leading to a mix of transverse and longitudinal $\gamma^*$ 
    polarizations. This may enhance the relative longitudinal  
    component depending on the angle and energy of the emitted gluon. 
    Dileptons produced by the NLO channels (mediated by gluons) 
    therefore have radically greater $\lambda_\theta^{\rm HX}$ 
    than their LO counterparts (see Fig. \ref{fig:cs_nlo}). 
    \item {\it Temperature effect}: 
    Dileptons produced during the earlier stages of 
    the fireball evolution have a larger $\lambda_\theta$, 
    because of a higher (effective) 
    temperature of the environment from which they are produced: 
    The dimensionless ratio which controls 
    $\lambda^{ }_\theta$ is $M/T$~\cite{Wu:2024vyc}, 
    and, from this perspective, 
    a large $T$ is tantamount to small $M$. 
\end{enumerate}

The gluon effect explains the fact that a smaller $\alpha$ results 
in a larger $\lambda^{ }_\theta$ 
(both the $M$ spectrum in Fig.~\ref{fig:lth_M_sf} 
and the IMR $\pT$-spectrum in Fig.~\ref{fig:lth_pT_IMR}), 
since a smaller $\alpha$ means the NLO channels in 
the pre-equilibrium stage are less suppressed. 
The temperature effect explains 
the difference between the cases with $\alpha=2$ and that 
without $\sf(T,\tau)$. 
Noticing that $\alpha=2$ means the LO and NLO channels are 
suppressed equally, 
the only difference between these two cases is whether 
the DPR in the pre-equilibrium stage is suppressed or not. 
With the pre-equilibrium suppression, 
dileptons from the earlier stage, although more polarized, 
have less weight when averaging with those created later.  
Therefore, the absence of $\sf(T,\tau)$ 
also increases the net $\lambda^{ }_\theta$, 
for both pre-equilibrium stage along and  
dileptons from the hydrodynamic stage.
\par 
In the collision system that we are reporting, 
it is observed in both 
Fig.~\ref{fig:lth_M_sf} and Fig.~\ref{fig:lth_pT_IMR} 
that for the pre-equilibrium stage along, the gluon effect wins 
over the temperature effect: 
$\alpha=1$ results in a greater polarization than not including 
$\sf(T,\tau)$. 
However, this ordering is reversed after mixing with the dileptons 
from the hydrodynamic stage. 

%
\begin{figure}[t]
\centering
\includegraphics[width=1.0\linewidth]{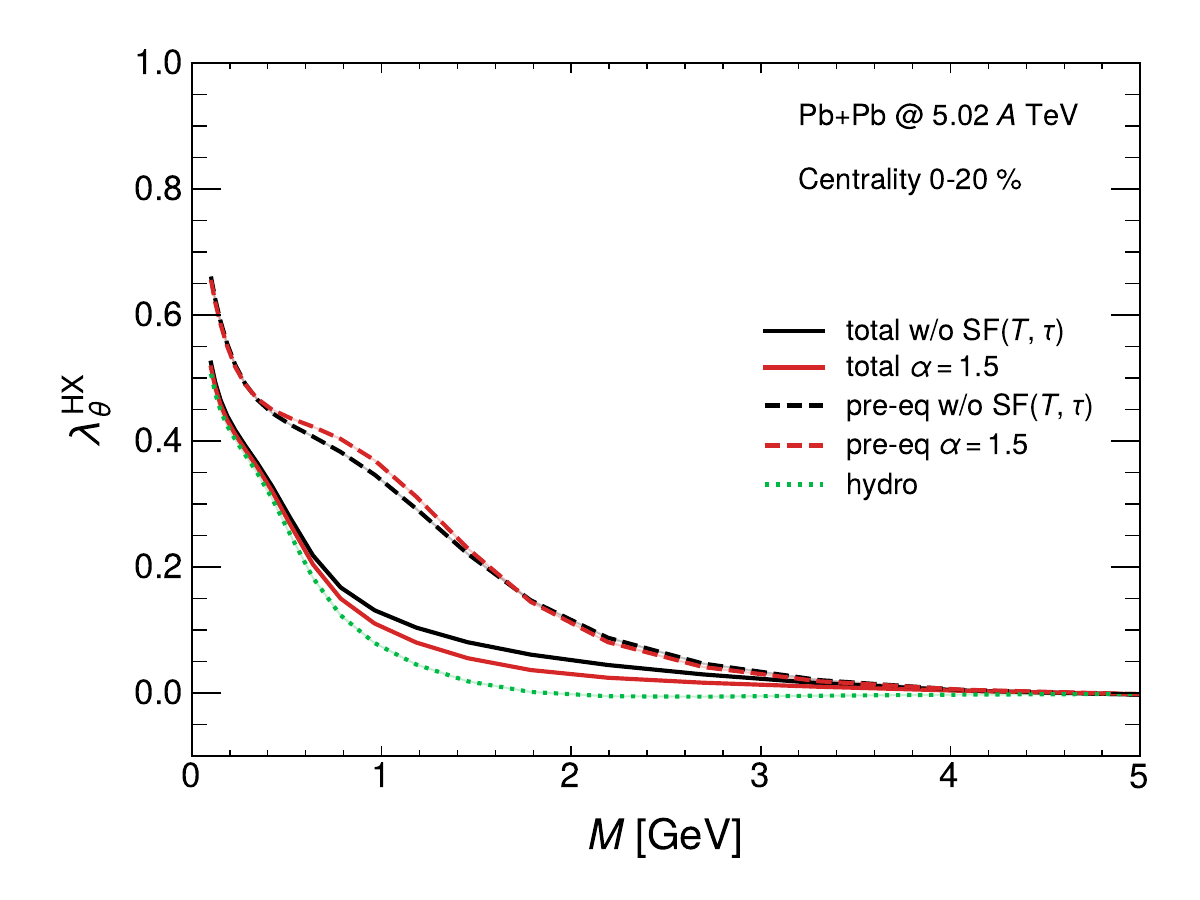}
\vspace{-6mm}
\caption{
  The invariant mass dependence of the 
  polarization coefficient $\lambda^{ }_\theta$ 
  for the pre-equilibrium dileptons (dashed lines),
  the total dileptons (solid lines) and 
  the hydrodynamic-stage dileptons (dotted green line). 
  Black lines represent the setting where no pre-equilibrium 
  quark suppression is introduced, 
  and red lines are for the case of the suppression index 
  $\alpha$ being the intermediate value of $1.5\,$.
}
\label{fig:lth_sf_M}
\end{figure}
%

%
\begin{figure}[t]
\centering
\includegraphics[width=1.0\linewidth]{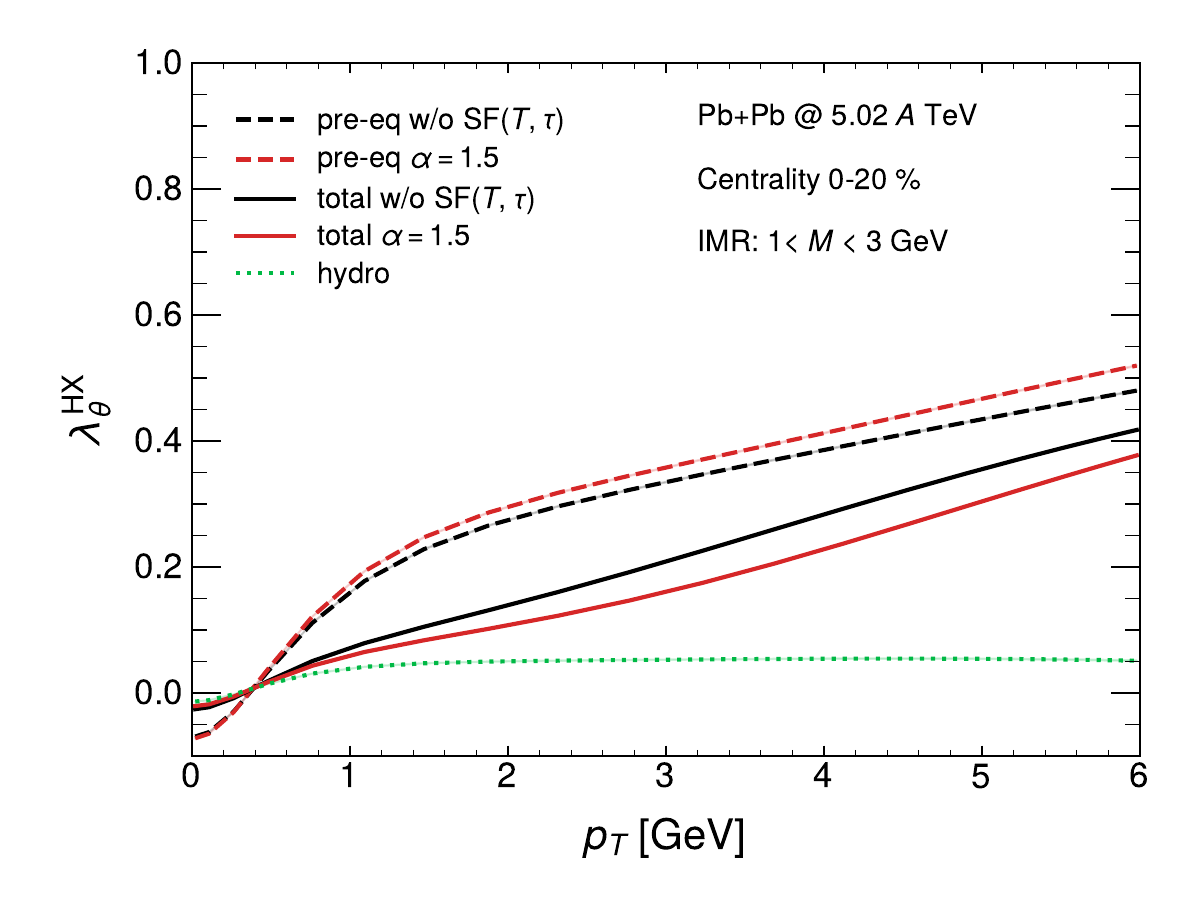}
\vspace{-6mm}
\caption{
  The coefficient $\lambda^{ }_\theta(\pT)$ for  
  pre-equilibrium dileptons (dashed lines), 
  the total dilepton yield (solid lines) and 
  the hydrodynamic-stage dileptons (dotted green line). 
  All calculations are performed for lepton pair in the IMR. 
  Black lines represent the setting where no pre-equilibrium 
  quark suppression is introduced, and red lines are for the case of 
  the suppression index $\alpha$ being the intermediate value of $1.5\,$.
}
\label{fig:lth_sf_pT_IMR}
\end{figure}
%

A similar behaviour is also observed for an intermediate 
value of $\alpha=1.5\,$, where the temperature effect makes both 
the total $\lambda^{ }_\theta(M)$ in Fig.~\ref{fig:lth_sf_M} and 
the IMR $\lambda^{ }_\theta(\pT)$ in Fig.~\ref{fig:lth_sf_pT_IMR} 
smaller than the case of no suppression factor. 

A net measurement of $\lambda^{ }_\theta$ in experiments 
is not sufficient to quantify 
the relative abundance of quarks and gluons in the pre-equilibrium stage. 
Nevertheless, one can still conclude that 
dileptons produced in the pre-equilibrium stage strongly contribute 
to the overall $\lambda_\theta^{\rm HX}$~\cite{Wu:2024vyc}, 
especially when considering their momentum dependence. 

\subsection{Adding Drell-Yan and thermal dileptons}
\label{sect:DY}

Drell–Yan dileptons in HICs are produced through hard partonic reactions 
at the first instant 
and carry characteristic polarization patterns set by perturbative QCD. 
The process is described within the framework of collinear factorization, 
whereby the short-distance scattering kernel is convoluted with 
non-perturbative 
nuclear parton distribution functions~\cite{Drell:1970wh}. 
As Drell-Yan dileptons predominate in the 
HMR,\footnote{%
  We still focus on $M \ll m^{ }_Z \,$, 
  so the intermediate $Z^0_{ }$ state does not need 
  to be considered.
} 
when combined with thermal dileptons emitted throughout the space–time 
evolution of the fireball, those contributions can influence the net 
observed polarization signals at large $M\,$.
\par
\begin{figure}[t]
\centering
\includegraphics[width=1.0\linewidth]{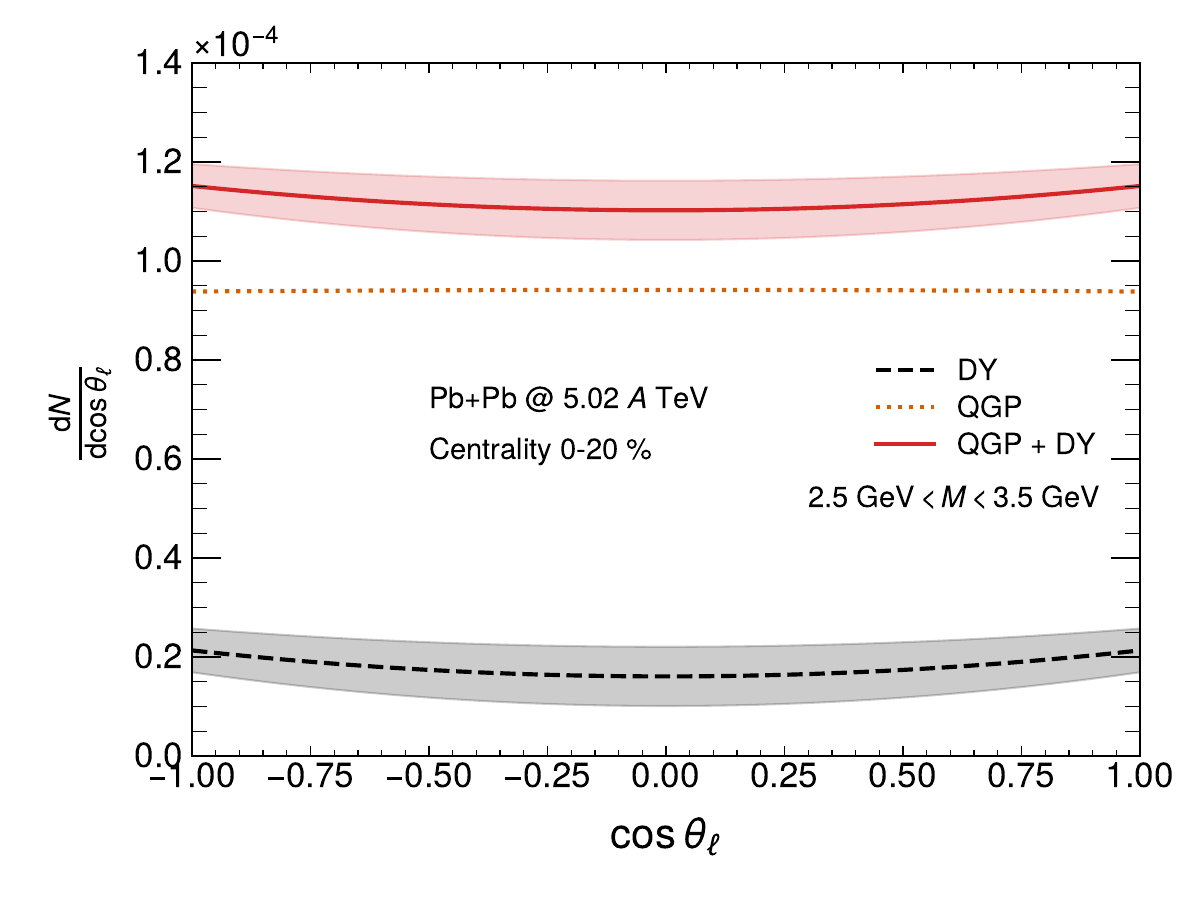}
\vspace{-6mm}
\caption{
  Angular distribution $\d N/\d\cos\theta_\ell$ of dileptons in 
  the invariant mass range $2.5~\mathrm{GeV} < M < 3.5~\mathrm{GeV}$. 
  The black dashed, orange dotted, and red solid lines show the NLO DY 
  (calculated using the DYTurbo package), 
  QGP, and the total (DY+QGP) contributions, respectively. 
  The shaded bands indicate the corresponding theoretical uncertainties 
  from the DY calculation, 
  obtained by varying the factorization and renormalization 
  scales in the range $\mu^{ }_{\rm F,R} = [\frac12, 2] \times M\,$. 
}
\label{fig:dNdcos}
\end{figure}

 
To allow a direct comparison with Ref.~\cite{Coquet:2023wjk}, 
we discuss polarization quadrupole moment in the CS frame. 
The polarization quadrupole moment is defined as
\begin{equation}
  \left\langle 
  \frac{3\cos^2\theta_\ell - 1}{2}
  \right\rangle 
  \; = \;
  \frac{3}{2}\frac{
  \raisebox{3mm}{$\displaystyle
    \int \d\Omega_\ell\, 
    \cos^2\theta_\ell\frac{\d N_{\ell\bar\ell}}{\d^4 K \d\Omega_\ell}
    $}
  }{
  \raisebox{-3mm}{$\displaystyle
    \int \d\Omega_\ell\, \frac{\d N_{\ell\bar\ell}}{\d^4 K d\Omega_\ell}
    $}
  } 
  - \frac{1}{2}
  \; , 
  \label{quadrupole def}
\end{equation}
where the differential yields can be  
evaluated in any virtual photon rest frame (including HX and CS). 
The relevant angular distribution is $\d N/\d\cos\theta_\ell\,$, 
which we show in Fig.~\ref{fig:dNdcos} at midrapidity for 
the invariant mass range $2.5 < M < 3.5~\mathrm{GeV}$. 
We calculate this angular dependence using the DYTurbo 
package~\cite{Camarda:2019zyx}, adapted for A+A collisions 
(see, e.g. Ref.~\cite{Wu:2024pba}). 

%
%

Substituting  the explicit form from Eq.~\eqref{eq:defpol} into the 
definition provided by Eq.~\eqref{quadrupole def} gives  
a direct relation between the polarization coefficient 
$\lambda^{ }_\theta$ and the quadrupole moment, namely
\begin{equation}\label{eq:q_lth}
  \left\langle \frac{3\cos^2\theta_\ell - 1}{2}\right\rangle 
  \; = \; 
  \frac{2}{15} \, 
  \frac{\lambda_\theta}{1+ \tfrac13 \lambda_\theta}
  \; .
\end{equation}
In a given frame, this equation allows us to translate from 
$\lambda^{ }_\theta$ to its corresponding quadrupole moment. 
Before averaging, $\frac12(3 \cos^2 \theta_\ell - 1)$ 
can range from $-\frac12$ to $1$, 
which would already restrict $\lambda_\theta$ in \eqref{eq:q_lth} 
to the semi-infinite ranges $(-\infty,-5]$ and $[-\frac53, + \infty)\,$. 
However, $(-\infty,-5]$ is not viable because $\lambda_\theta \geq -1$ 
for a positive yield in Eq.~\eqref{eq:defpol}. 
Typically $\lambda_\theta$ also cannot be too large, 
although the precise range depends on the frame and the averaging 
procedure, cf.~Appendix~\ref{app:qm}. 
For $|\lambda_\theta| \lesssim 1$, the mapping is almost linear, 
so one does not expect a qualitative difference between these 
two languages. 
An example of such translation is given in the Appendix~\ref{app:qm}.
\par 
Figure~\ref{fig:Qmix} shows that the quadrupole contributions 
of Drell-Yan and thermal dileptons (calculated at NLO) have 
the opposite sign.  
The sizable positive quadrupole momentum of Drell-Yan dileptons 
is also seen in Ref.~\cite{Coquet:2023wjk}.\footnote{%
  Note also that $\lambda^{\rm CS}(\theta) \equiv$ 1 at LO for Drell-Yan.
} 
Drell-Yan dileptons start to manifest themselves in the 
invariant mass spectra $\d N/\d M$ around $M\gtrsim 2~{\rm GeV}$, 
below which point collinear factorization becomes unreliable.
Drell-Yan
eventually wins over the thermal counterpart for $M>4.5~{\rm GeV}$. 
In most of the IMR, despite their yield being roughly an 
order-of-magnitude lower than that of thermal dileptons, 
the Drell-Yan polarization signature is considerable and becomes 
even more pronounced for larger $M$, as expected. 
Let us mention that in Ref.~\cite{Coquet:2023wjk}, the pre-equilibrium 
quadrupole moment was found to be around $\sim -0.04\,$, which has the 
same sign as our result but is significantly larger in magnitude. 
In that calculation, which was at LO, the behaviour was explained
from the anisotropy developed in the 
colliding quark and antiquark momentum distributions. 
Our results are at NLO, where the phase space is more complicated, 
but we have simplified the pre-equilibrium treatment 
by assuming isotropic thermal-like distributions 
(see footnote~\ref{foot:distortions}). 

%
\begin{figure}[t]
\centering
\includegraphics[width=1.0\linewidth]{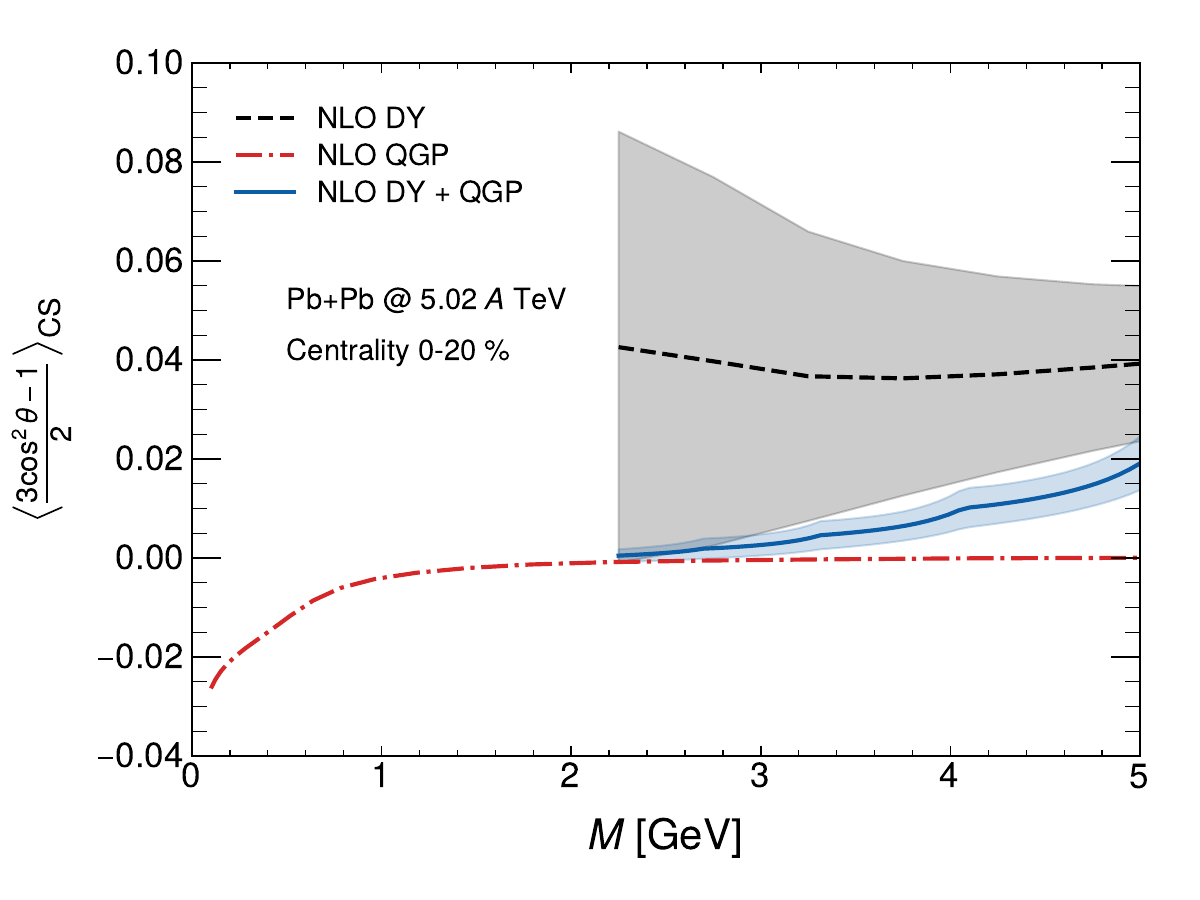}
\vspace{-6mm}
\caption{
  Invariant mass dependence of the CS frame quadrupole momentum, 
  as defined in Eq.~\eqref{quadrupole def}. 
  The black dashed line shows the NLO Drell–Yan (DY) contribution, 
  the red dash-dotted line corresponds to the NLO QGP thermal contribution, 
  and the blue solid line represents their sum. 
  The shaded bands represent the uncertainty, as in Fig.~\ref{fig:dNdcos}. 
}
\label{fig:Qmix}
\end{figure}
%

\subsection{Dimuon polarization}

So far, we have only discussed dielectron polarization where 
$m_e = 0.511~{\rm MeV}$ and thus the kinematic threshold 
for $M \simeq m_e$, described by the factor $B$ in Eq.~\eqref{eq:d4rate}, 
is of little practical relevance. 
Hence the parameter $\xi = m_\ell^2/M^2$, appearing in particular 
in Eqs.~\eqref{eq:ltheta} to \eqref{eq:lperpthph}, doesn't 
manifest itself in our predictions. 
If heavier dileptons, for instance, $\uu$ pairs, are considered, 
the approximation appearing in Eq.~\eqref{eq:lth_approx} 
will be modified by 
$\xi = m_\mu^2/M^2$ with $m_\mu = 0.106{\rm~GeV}$.  
Previous experimental measurements of dilepton polarization 
also looked at $\uu$~\cite{NA60:2008ctj} instead of $\ee$. 
In this section, we will briefly present and discuss 
our predictions on dimuon polarization. 
\par
One important corollary from Eq.~\eqref{eq:ltheta} on this topic 
is that $\lambda_\theta^\ee$ and $\lambda_\theta^\uu$ are strictly
related to one another.  
\emph{Regardless of the local flow velocity} $\vec u$ and 
independent of the choice of $\gamma^*$ rest frame, 
if one makes the approximation $m_e \simeq 0$, 
for every point in the dilepton phase space, 
the following identity holds
\begin{equation}\label{eq:uu_from_ee}
  \lambda_\theta^\uu 
  \; = \; 
  \frac{
    (1-4m_\mu^2/M^2) \, \lambda_\theta^\ee
  }{
    1 + (2m_\mu^2/M^2)(1+\lambda_\theta^\ee)
  }
  \; .
\end{equation}

%
\begin{figure}[t]
\centering
\includegraphics[width=1.0\linewidth]{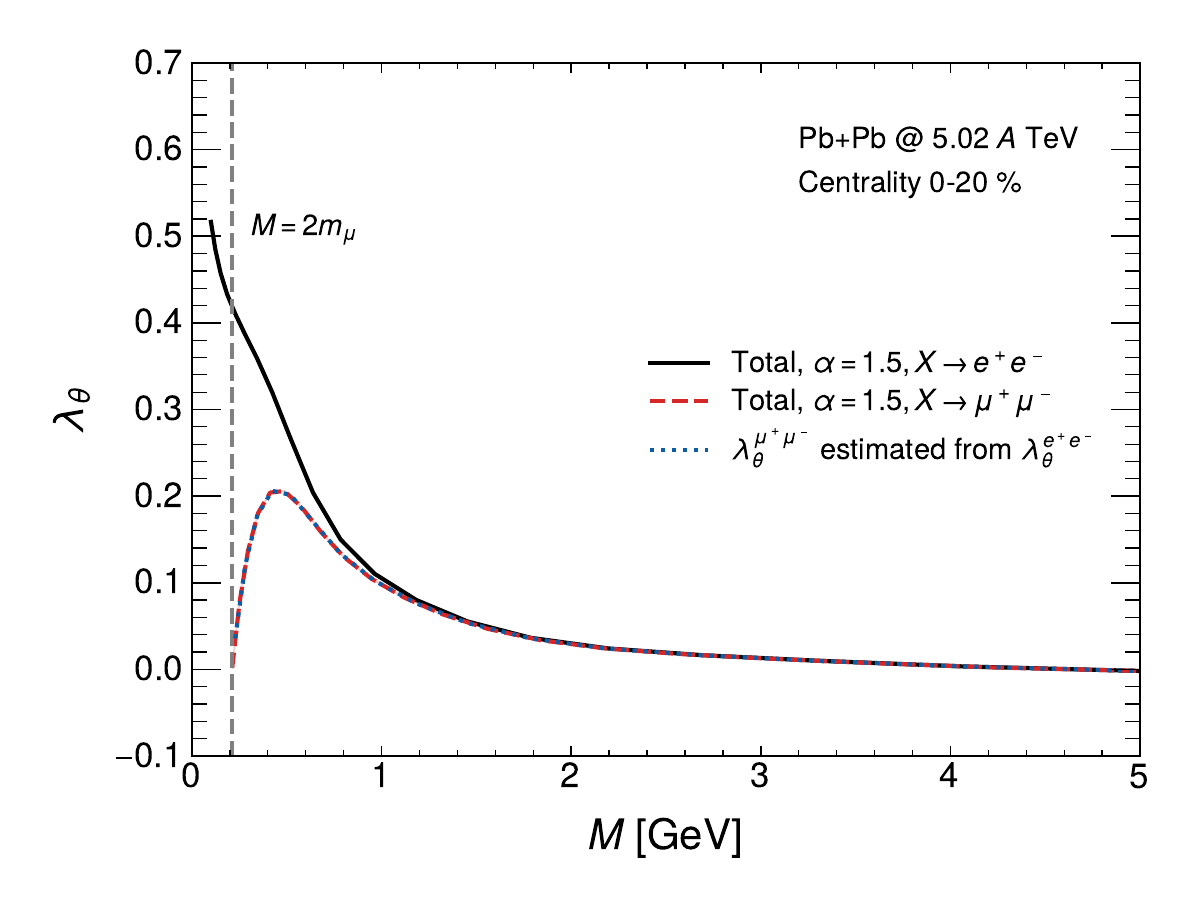}
\vspace{-6mm}
\caption{
  Comparing the invariant mass dependence of the HX frame 
  polarization coefficient $\lambda_\theta^{ }$ of 
  thermal dielectrons (black solid line) and dimuons (red dashed line). 
  The blue dotted line shows inferring dimuon 
  $\lambda_\theta^{ } (M)$ from the dielectron line using 
  Eq.~\eqref{eq:uu_from_ee}, and it overlaps exactly with 
  the red dashed line, which is the direct model prediction of 
  $\lambda_\theta^\uu (M)\,$.
}
\label{fig:lth_sf_M_est}
\end{figure}
%

Because of a cancellation of the denominator 
and a factor in the weight function (cf.~Eq.~\eqref{eq:defN}), 
this identity will be preserved 
even after averaging over the QGP space-time history 
or the $\gamma^*$ phase space. 
In Fig.~\ref{fig:lth_sf_M_est}, we verify this fact by presenting 
the invariant mass $\lambda_\theta$ spectrum for $\uu$ and $\ee$ 
directly calculated by our model, 
along with $\lambda_\theta^\uu\!\!(M)$ 
using Eq.~\eqref{eq:uu_from_ee}. 
A perfect match is indeed found. This can be well understood in 
terms of the underlying spectral functions: The medium affects the 
electromagnetic thermal correlator in 
Eq.~\eqref{eq:d6rate} via $\rho^{\mu\nu}$, while the type of  
lepton only affects the vacuum matrix element $L^{ }_{\mu\nu}\,$. 
\par
In practice, our findings suggest that for thermal dileptons, 
$\ee$ and $\uu$ measurements are not independent and can always 
be related by a simple identity Eq.~\eqref{eq:uu_from_ee}. 
Additional model predictions for $\uu$ spectra can be found 
in Appendix~\ref{app:dimuon}.

%
\section{Summary}\label{sect:summary}

In this paper, we present a comprehensive theoretical study 
of dilepton polarization coefficients based on a perturbative 
description of underlying partonic QGP rates. 
Using our framework, which combines NLO photon spectral functions  
with a state-of-art multi-stage hydrodynamic simulation 
of the evolving fireball, we discuss various aspects 
of dilepton polarization,  focusing on the IMR.
\par
Polarization coefficients are calculated in both the HX and CS frames, 
in the presence of hydrodynamic evolution. 
For a static medium, the values of $\lambda^{ }_\theta$ in these two frames 
are directly correlated, while for a realistic QGP, 
the CS frame polarizations are found to be more sensitive to flow. 
We also confirm both theoretically and numerically that 
$\tilde \lambda$, defined in Eq.~\eqref{eq:invlambda} for a point source,
is a frame-independent observable 
for dileptons accumulated over the entire fireball, 
see Fig.~\ref{fig:lambda_NLO}.
\par
We find that the polarization of dileptons in distinct mass 
regions is dominated by different sources. 
In the IMR, the polarization spectrum 
is sensitive to 
thermal dileptons emitted during the hydrodynamic stage 
of the evolution. 
Furthermore, our phenomenological treatment of dilepton production 
during the pre-equilibrium stage shows that the momentum dependence 
in the IMR dilepton is especially sensitive to the presence of 
pre-equilibrium dynamics. 
In contrast, Drell-Yan dileptons, 
produced in the initial hard scatterings of the collision, 
dominate the HMR.
\par 
Lastly, a strict one-to-one correspondence of dielectron polarization 
and dimuon polarization is inferred from our theoretical framework, 
and is further confirmed numerically by our model calculation. 
This correspondence should guide theoretical calculations and 
experimental measurements which respectively focus 
on dielectrons and dimuons. 
\par
We have focused on lepton pair production in Pb+Pb collisions at 
$\sqrt{s_{\rm NN}} = 5.02$~TeV, i.e. in conditions typical of the LHC. 
Note that there already exist polarization measurements in 
heavy-ion collisions  made by 
NA60~\cite{NA60:2008iqj} and 
HADES~\cite{HADES:2011nqx} at lower energies. 
Future work which will include the LMR spectral densities 
will be devoted to those results. 

\section{Acknowledgements}
We thank Maurice Coquet for clarifications regarding 
Ref.~\cite{Coquet:2023wjk}. 
This work was supported in part by the Natural Sciences 
and Engineering Research Council of Canada  (NSERC) 
[SAPIN-2026-00047 (C.G.) and SAPIN-2024-00026 (S.J.)], 
and in part (G.J.) by 
the Agence Nationale de la Recherche (France), 
under grant ANR-22-CE31-0018 (AUTOTHERM). 
Computations were made on the B\'eluga and 
the Rorqual super-computer systems managed by Calcul Qu\'ebec and 
by the Digital Research Alliance of Canada.

%
\appendix

%
\section{Degrees of freedom in polarization\\[1mm] coefficients}
\label{app:dof}

According to Eqs.~\eqref{eq:ltheta} to \eqref{eq:lperpthph}, 
for thermal dileptons 
the five polarization coefficients are fully determined 
by $\rho_{_{\rm T}}/\rho_{_{\rm L}}$ and $\vec u_*/|\vec u_*|\,$, 
which contain three degrees of freedom in total 
(one from $\rho_{_{\rm T}}/\rho_{_{\rm L}}$, 
and the unit three-vector $\vec u_*/|\vec u_*|$ contains two). 
It is therefore possible to construct a clean combination 
of these polarization coefficients with no
$\vec u_*$ dependence at all. 
For example, by noting that 
$\chi_z = \chi_{xz}\chi_{yz}/\chi_{xy}$, 
$\chi_z$ can be obtained by 
$
  \mu 
  \equiv 
  \lambda_{\theta\phi} \lambda_{\theta\phi}^\perp
  /\lambda_\phi^\perp
$. 
Substituting this into Eq.~\eqref{eq:ltheta} gives an expression 
for $\rho_{_{\rm T}}/\rho_{_{\rm L}}$ involving only 
the polarization coefficients
\begin{equation}\label{eq:rtl}
  \frac{\rho_{_{\rm T}}}{\rho_{_{\rm L}}} 
  \; = \;
  \frac{
  1-4\xi  - (1+4\xi)\lambda_\theta + 2\mu(1+2\xi)
  }{
  (3+4\xi) \lambda_\theta + (1-4\xi)-4\mu(1+2\xi)]
  }.
\end{equation}
If one takes the $m_\ell \simeq 0$ approximation 
(so $\xi \simeq 0$), as appropriate for dielectrons, 
Eq.~\eqref{eq:rtl} further reduces to
\begin{equation}\label{eq:rtl2}
  \frac{\rho_{_{\rm T}}}{\rho_{_{\rm L}}} 
  \; \simeq \;
  \frac{(1- \lambda_\theta)\lambda_\phi^\perp
  +
  2\lambda_{\theta\phi}\lambda_{\theta\phi}^\perp  
  }{
  (1+3 \lambda_\theta)\lambda_\phi^\perp 
  -
  4 \lambda_{\theta\phi}\lambda_{\theta\phi}^\perp 
  }
  \; .
\end{equation}
For the QGP created in a symmetric collision, 
the cross terms $\chi_{ij} = u_*^i u_*^j/\vec u_*^2$ ($i\not=j$) 
are small after averaging over the entire fireball and the 
dilepton's direction $\hat k\,$. 
This causes both the numerator and the denominator 
of Eq.~\eqref{eq:rtl2} to be very small. 
In our model calculation, the magnitudes of 
$\lambda_{\theta\phi}\,$,
$\lambda_{\phi}^\perp$ 
and 
$\lambda_{\theta\phi}^\perp$ 
are all $\lesssim 10^{-20}$. 
With experimental uncertainties in mind, 
the right hand side of Eq.~\eqref{eq:rtl2} therefore behaves 
like an undetermined ``$0/0$'' ratio, rendering it 
an impractical estimator 
of polarization through the value of $\rho_{\rm _T}/\rho_{\rm _L}$. 

%
\section{Polarization coefficients and\\[1mm] quadrupole moments}
\label{app:qm}

\begin{figure}[t]
\centering
\includegraphics[width=1.0\linewidth]{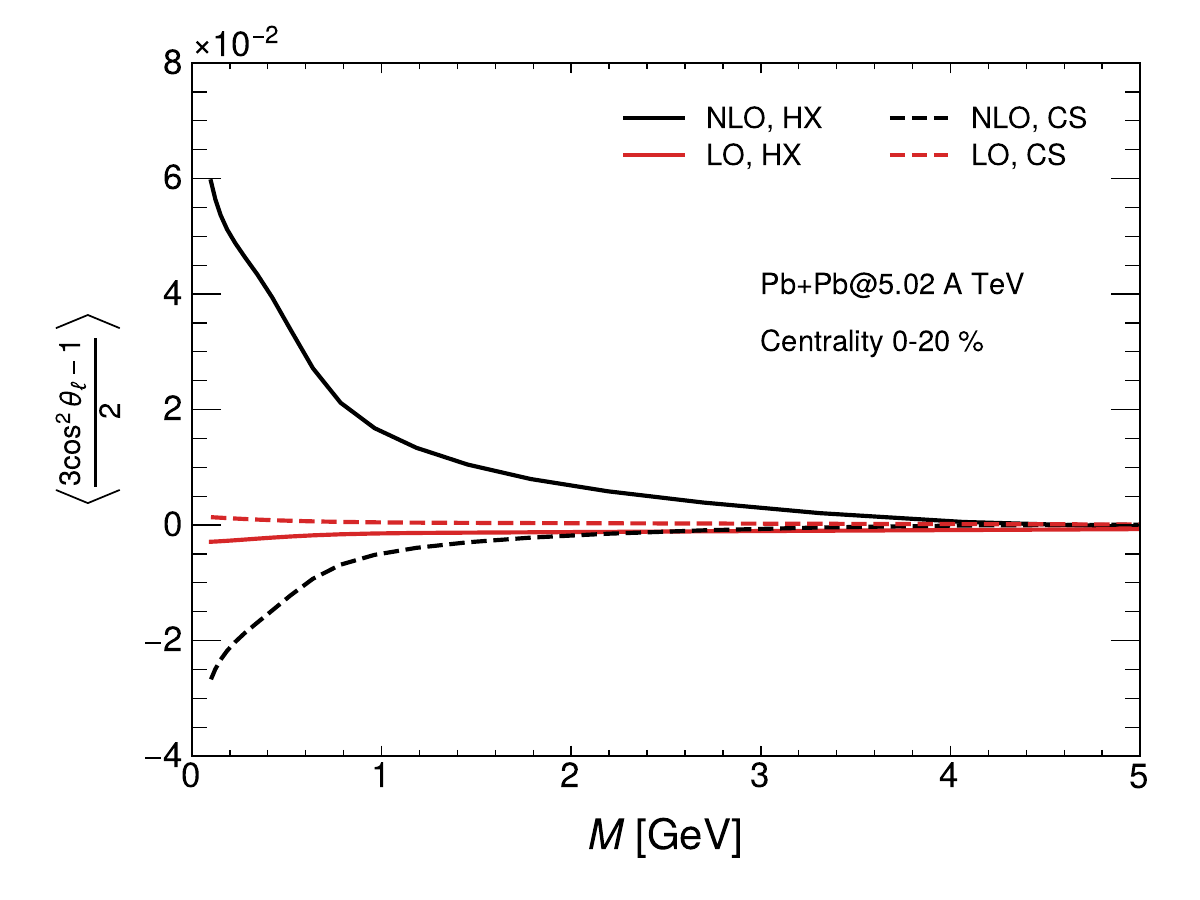}
\vspace{-6mm}
\caption{
  LO (red lines) and NLO (black lines) predictions of 
  the quadrupole moment 
  $\left\langle \frac{3\cos^2\theta_\ell - 1}{2}\right\rangle$ 
  invariant mass spectrum in the HX (solid lines) 
  and CS (dashed lines) frames, 
  related to Fig.~\ref{fig:cs_nlo} by Eq.~\eqref{eq:q_lth}. 
}
\label{fig:quadrupole_frame}
\end{figure}

To better demonstrate the equivalence between the 
$\lambda^{ }_\theta$ and quadrupole moment observables, 
we show the quadrupole moment version of Fig.~\ref{fig:cs_nlo} 
in Fig.~\ref{fig:quadrupole_frame}, 
where all the curves follow the same trend in both figures.

%
%

In Ref.~\cite{Coquet:2023wjk}, 
the $z$-axis is defined according to the Collins-Soper prescription.  
The angle $\theta$ is the polar angle made by one of the final leptons, 
say $\ell\,$, 
in the centre-of-mass frame (i.e. the virtual photon rest frame). 
Those authors study the LO process, $q \bar q \to \ell \bar \ell\,$, 
in this frame for 
which it is well-known that the  angular dependence is 
${\rm d}\sigma/{\rm d}\Omega \sim 1 + (\hat p_q \cdot \hat p_\ell)^2$ 
where $\hat p_i$ is the direction of colliding particle $i\,$. 
The $x$-axis may be chosen so that 
\begin{equation}
  \vec{p}_\ell 
  \; = \; 
  E 
  (\, \sin \theta \,,\, 0 \,,\, \cos \theta \,) 
  \, ,
\end{equation}
with $\vec p_{\bar \ell} = - \vec p_\ell\,$ 
(final leptons are back-to-back). 
The initial quarks can have any direction, although 
two cases are discussed in Ref.~\cite{Coquet:2023wjk}, 
we consider the general situation:
\begin{equation}
  \vec{p}_q 
  \; = \; 
  E 
  (\, 
    \cos \alpha \sin \beta \,,\, 
    \sin \alpha \sin \beta \,,\, 
    \cos \beta
  \,) 
  \, , 
\end{equation}
with $\vec p_{\bar q} = - \vec p_q\,$. 
For this choice of coordinate system and these angles, 
the cross section goes as
\begin{align}
  \frac{ {\rm d}\sigma }{ {\rm d}\Omega } 
  &\sim 
  1 + \cos^2 \alpha\, \sin^2 \beta + 
  (\cos^2 \beta - \cos^2 \alpha\, \sin^2 \beta) \cos^2 \theta 
  \nonumber\\
  & + 2 \cos \alpha \, 
  \sin \beta \, \cos \beta \, 
  \sin \theta \, \cos \theta 
  \, .
  \label{LO cross section}
\end{align}
The second line above will always drop out when computing the 
quadrupole moment from  Eq.~\eqref{quadrupole def}. 
Reference~\cite{Coquet:2023wjk} then averages over the 
initial azimuthal angle $\alpha\,$, 
i.e. replacing $\sin^2 \alpha$ and $\cos^2 \alpha$ by $1/2$ in 
Eq.~\eqref{LO cross section}. 
One may then read-off:
\begin{equation}
  \lambda^{{\rm CS}, q\bar q \to \ell \bar \ell}_\theta(\beta) 
  = 
  \frac{1 - \frac32 \sin^2 \beta}{1 + \frac12 \sin^2 \beta} 
  \, .
\end{equation}
One may easily check that $- 1/3 \leq \lambda_\theta(\beta) \leq 1\,$, 
with the lower bound attained for $\beta = \pm \pi/2$ 
and the upper bound attained for $\beta = 0, \pi\,$. 
These latter two extremes are precisely the ones 
considered by \cite{Coquet:2023wjk}, respectively 
transverse initial quarks 
($\lambda_\theta = - \frac13$, giving a quadrupole moment of $- \frac1{20}$)
and initial quarks parallel with $z$ 
($\lambda_\theta = 1$, giving a quadrupole moment of $\frac1{10}$).

If both $\alpha$ and $\beta$ are averaged over, one trivially 
obtains $\lambda_\theta = 0\,$, which may seem like it should 
represent the LO case for a thermal medium. 
However, for such systems, 
there is also a flow velocity $\vec u$ as well as 
additional processes like $q \to q \gamma^*$ 
(where the final quark gets reabsorbed by the medium) 
which are encoded by the spectral functions.

%
\section{Dimuon polarization spectra}
\label{app:dimuon}

\begin{figure}[t]
\centering
\includegraphics[width=1.0\linewidth]{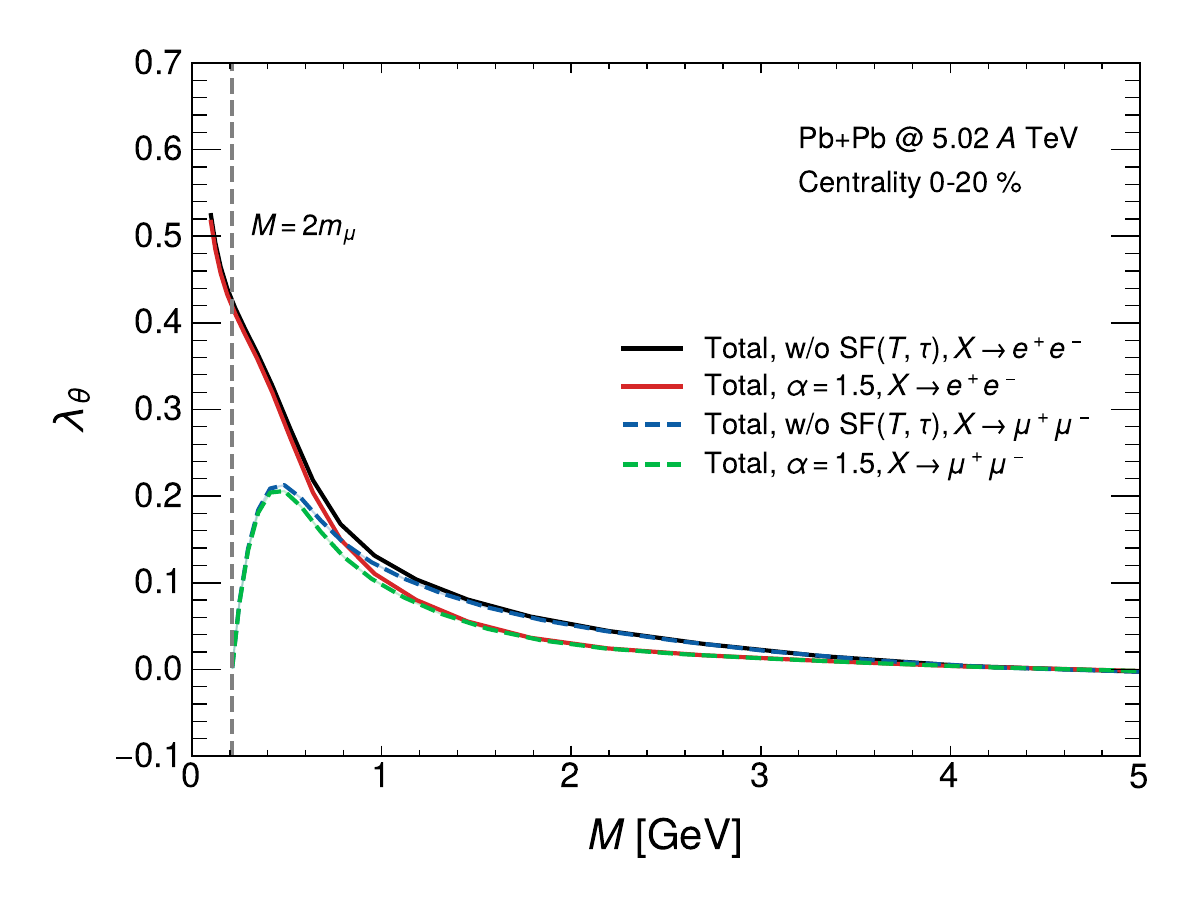}
\vspace{-6mm}
\caption{
  The invariant mass dependence of the helicity-frame polarization 
  coefficient $\lambda^{ }_\theta\,$, 
  integrated over $\pT$. 
  Solid (dashed) lines illustrate the dielectron (dimuon) results, 
  with and without the suppression factor (SF) modification as indicated. 
  The vertical dashed line marks the kinematic threshold 
  to produce the dimuon pair, namely $M = 2m_\mu$.
}
\label{fig:lth_sf_M_mu}
\end{figure}

For completeness, in this appendix, we briefly conclude how our 
earlier findings on pre-equilibrium dileptons will change 
if dimuon observables are considered instead of dielectrons. 
Figure~\ref{fig:lth_sf_M_mu} compares the dimuon and dielectron 
HX-frame $\lambda^{ }_\theta(M)$, for the scenarios of 
no suppression factor and using the suppression factor with 
$\alpha=1.5$ in Eq.~\eqref{eq:dr_pre-eq}. 
In addition to a clear threshold effect at $M = 2m_\mu$, 
the dimuon $\lambda^{ }_\theta(M)$ has a smaller magnitude than 
its dielectron counterpart in the LMR region 
($M\lesssim 1~{\rm GeV}$), but follows the same trend. 
The differences in $\lambda^{ }_\theta(M)$ brought about by $\sf(T,\tau)$, 
discussed in Sec.~\ref{ssect:pre-eq}, are essentially 
the same for dielectrons and dimuons.
\par
\begin{figure}[t]
\centering
\includegraphics[width=1.0\linewidth]{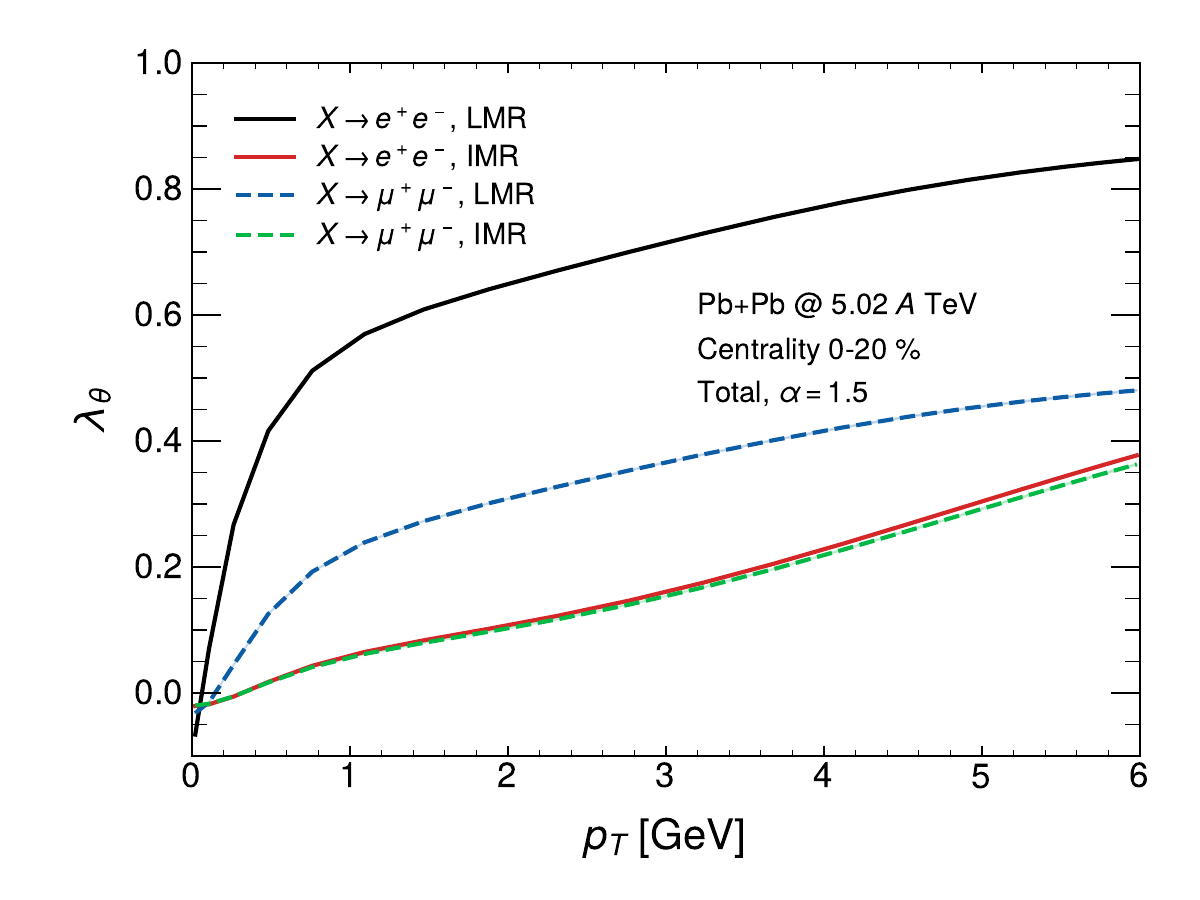}
\vspace{-6mm}
\caption{
  The $\pT$-dependence of the helicity-frame polarization 
  coefficient $\lambda_\theta\,$, 
  integrated over $M$ in different mass ranges. 
  Solid (dashed) lines illustrate the dielectron (dimuon) results, 
  with and without the suppression factor (SF) 
  modification as indicated.
}
\label{fig:lth_sf_pT_mu}
\end{figure}

We further examine the difference in $\lambda^{ }_\theta(\pT)$ 
for dielectrons and dimuons in Fig.~\ref{fig:lth_sf_pT_mu}. 
As observed in Fig.~\ref{fig:lth_sf_M_mu}, 
the dielectron $\lambda^{ }_\theta$ is significantly greater 
than that of dimuons in the LMR, 
while little difference is found in the IMR. 

\bibliography{references}

\end{document}